\documentstyle[epsf,epsfig,rotate]{mn2e}

\newcommand\bb[1] {   \mbox{\boldmath{$#1$}}  }
\newcommand\bcdot{\bb{\cdot}}
\newcommand{\nab}{\mbox{\boldmath $\nabla$}}
\newcommand{\tim}{\mbox{\boldmath $\times$}}
\newcommand{\xib}{\mbox{\boldmath $\xi$}}
\newcommand{\beq}{ \begin{equation} }
\newcommand{\eeq}{ \end{equation} }

\def\spose#1{\hbox to 0pt{#1\hss}}
\def\ltsim{\mathrel{\spose{\lower.5ex\hbox{$\mathchar"218$}}
         \raise.4ex\hbox{$\mathchar"13C$}}}

\onecolumn

\title[Numerical simulations of planetary migration in magnetized discs]
{Numerical simulations of type I planetary migration in nonturbulent 
magnetized discs}

\author[Fromang, Terquem \& Nelson]
{S\'ebastien Fromang$^1$, Caroline Terquem$^{2,3,4}$ and Richard P. Nelson$^{1}$ \\
$^1$ Astronomy Unit, Queen Mary, University of London, Mile End Road, 
London E1 4NS \\
$^2$ Institut d'Astrophysique de Paris, UMR7095 CNRS,
Universit\'e Pierre \& Marie Curie--Paris~6, 98bis boulevard Arago,
75014 Paris, France \\
$^3$ Universit\'e Denis Diderot--Paris 7, 2 Place Jussieu, 75251
Paris Cedex 5, France \\
$^4$ Institut Universitaire de France }

\date{Accepted.
      Received;
      in original form }

\pubyear{}

\begin{document} 

\maketitle

\begin{abstract} 
Using 2D MHD numerical simulations performed with two different finite
difference Eulerian codes, we analyze the effect that a toroidal
magnetic field has on low mass planet migration in nonturbulent
protoplanetary discs.  The presence of the magnetic field modifies the
waves that can propagate in the disc. In agreement with a recent
linear analysis (Terquem 2003), we find that two magnetic resonances
develop on both sides of the planet orbit,
which contribute to a significant global torque.  In order to measure
the torque exerted by the disc on the planet, we perform
simulations in which the latter is either fixed on a circular orbit or
allowed to migrate. For a $5$~earth mass planet, when the ratio
$\beta$ between the square of the sound speed and that of the Alfven
speed at the location of the planet is equal to $2$, we find inward
migration when the magnetic field $B_{\phi}$ is uniform in the disc,
reduced migration when $B_{\phi}$ decreases as $r^{-1}$ and outward
migration when $B_{\phi}$ decreases as $r^{-2}$. These results are in
agreement with predictions from the linear analysis.  Taken as a
whole, our results confirm that even a subthermal stable field can
stop inward migration of an earth--like planet.
\end{abstract}  


\section{Introduction} 

According to our present knowledge, two main scenarios are favored to
describe giant planet formation. In the first, planets form via
massive disc fragmentation resulting from strong gravitational
instabilities (see for example Boss, 1998). The second scenario is the
so--called core--accretion mechanism (Pollack et al. 1996), according
to which small bodies accumulate to form cores of a few earth masses
that eventually accrete an envelope which leads to a gas giant
planet. In both cases, young planets are supposed to be embedded in
their parent accretion disc during the first few million years of
their evolution.

In the protoplanetary disc, the protoplanet excites density waves at
the Lindblad resonances that propagate on both sides of its orbit
(Goldreich \& Tremaine 1979).  In the linear regime, the net torque
these waves exert on the planet is negative, which causes the planet
to migrate inward in the disc (Ward 1997).  In the nonlinear regime,
the planet opens up a gap in the disc and is then locked into the disc
viscous evolution, which also induces inward migration (Lin and
Papaloizou 1993 and references therein).  This is the
mechanism that is believed to be at the origin of the population of
giant exoplanets whose period is observed to be smaller than $10$ days
(Udry et al. 2003).

For low mass planets like the earth, the migration timescale is of the
order of $10^5$ years (Tanaka et al. 2002), much smaller than the
disc's lifetime, which is believed to be roughly $10^7$ years. A
physical mechanism is needed to halt this inward migration and account
for the existence of planetary systems. To date, a few possibilities
have been investigated to explain why Jupiter--like planets would stop
their inward migration (Trilling et al, 1998; Lecar \& Sasselov, 2003;
Matsuyama et al, 2003). For lower mass planets, a mechanism involving
a toroidal magnetic field has recently been put forward by Terquem
(2003, hereafter T03). The effect of the magnetic field is to allow
new waves to propagate in the disc, which have an effect on the torque
that is exerted on the planet. Linearizing the MHD equations, Terquem
(2003) indeed showed that outward migration can be induced on the
planet when the magnetic field is rapidly decreasing with radius. The
goal of the present study is to extend this analysis outside of the
linear approximation by performing 2D numerical simulations of a
planet embedded in a razor--thin accretion disc in the presence of a
toroidal magnetic field. Our aim is to study the magnetic resonances
identified in T03 and to measure the torques exerted by the disc on
the planet.

In nature, such a configuration, consisting of a toroidal field
embedded in a Keplerian disc, is expected to be unstable because of
the magnetorotational instability and lead to MHD turbulence (Balbus
\& Hawley 1998). In our 2D simulations, however, the toroidal magnetic
field is stable. The absence of MHD turbulence makes it easier to
analyze the properties of the waves that propagate in the disc and to
accurately calculate the torque exerted by the disc on the planet. The
question of how MHD turbulence will affect the results presented here
is an important one. It is beyond the scope of this preliminary work,
but needs to be addressed in future studies. We note that there are
existing simulations on low--mass planet migration in MHD turbulent
discs (Nelson \& Papaloizou 2004), but they lack the resolution
required to study the magnetic resonances. Indeed, the size of the cells 
in tese simulations is only half the distance between the planet and the expected 
location of the resonances (which lie very close to the planet for the 
low field generated by the turbulence) and also the smoothing 
length used for the gravitational potential is much larger than this 
distance.

The plan of the paper is as follows. In section~{2}, we develop a
simple analytical model showing the basic properties of wave 
propagation in the presence of a toroidal magnetic field. In
section~{3} we describe the numerical simulations with which we
analyze the magnetic resonances. Our results, performed with two
completely independent numerical codes, include cases for which the
planet is fixed on a circular orbit and cases for which it is allowed
to migrate in the disc.  Finally, we discuss the results of our work
in section~{4}.

\section{Wave propagation in a magnetized shear flow}

In a non magnetized disc, the linear perturbation exerted by a small
mass protoplanet propagates as density waves outside the Lindblad
resonances and is evanescent inside these resonances, in the
corotation region.  The protoplanet exerts a torque on the density
waves, which, together with the torque exerted at corotation, is
responsible for the exchange of angular momentum between the disc's
rotation and the planet's orbital motion.  The angular momentum
carried by the density waves is advected through the disc, and
possibly transferred to the disc if the waves are dissipated, while
the torque exterted at corotation is transferred directly to the disc
material (Goldreich \& Tremaine 1979).

The interaction between the planet and the disc inside/outside its
orbit beyond the Lindblad resonances leads to a negative/positive
torque on the disc, and therefore to a gain/loss of angular momentum
for the planet.  In the linear regime, because in a (even uniform)
Keplerian disc the outer Lindblad resonances are slightly closer to
the planet than the inner Lindblad resonances, the interaction with
the outer parts of the disc leads to a larger Lindblad torque than
that with the inner parts (Ward 1986, 1997).  Therefore, the net
Lindblad torque exerted by the planet causes it to lose angular
momentum and to move inward relative to the gas (type~I migration).

Wave propagation in a Keplerian disc containing a (stable) toroidal
magnetic field and an embedded low mass planet was studied in T03.  It
was found that all fluid perturbations are singular at the so--called
{\em magnetic resonances}, where the Doppler shifted frequency of the
perturbation matches that of a slow MHD wave propagating along the
field lines.  These lie on both sides of the corotation radius.  Waves
propagate outside the Lindblad resonances, and also in a restricted
region around the magnetic resonances.  It was found that the torque
exerted in the vicinity of the magnetic resonances tends to dominate
the disc response when the magnetic field is large enough.  This
torque, like the Lindblad torque, is negative inside the planet's
orbit and positive outside the orbit.  Therefore, if the magnetic
field decreases fast enough with radius, the outer magnetic resonance
becomes less important (it disappears altogether when there is no
magnetic field outside the planet's orbit) and the total torque
becomes negative, dominated by the inner magnetic resonance.  This
corresponds to a positive torque on the planet, which leads to outward
migration.

It if important to know where the waves can propagate to understand
how angular momentum is transferred between the planet's orbital
motion and the disc's rotation.  However, the full problem (with
magnetic field) is too complex to allow for the characterization of
the waves in the vicinity of the planet.  Therefore, here we adopt a
simpler toy model which will enable us to get a better understanding
of the dynamics around the planet's orbit, and to interpret the
numerical simulations that are presented in this paper. 

\subsection{Basic equations}
\label{mhd equations}

We consider a two dimensional flow which is described by the equation
of motion:

\begin{equation}
\Sigma \left[ \frac{\partial {\bf v}}{\partial t} + 
\left( {\bf v} \cdot \nab \right) 
{\bf v} \right] = - \nab P +  {\bf F} ,
\label{motion}
\end{equation}

\noindent the equation of continuity:

\begin{equation}
\frac{\partial \Sigma}{\partial t} + \nab \cdot
\left( \Sigma {\bf v} \right) = 0 ,
\label{mass}
\end{equation}

\noindent and the induction equation in the ideal MHD approximation:

\begin{equation}
\frac{\partial {\bf B}}{\partial t} = 
\nab \tim \left( {\bf v} \tim {\bf B} \right),
\label{induction}
\end{equation}

\noindent where 

\begin{equation}
{\bf F} = \frac{1}{\mu_0} \left( \nab \tim {\bf B} \right) \tim {\bf
B}
\label{lorentz}
\end{equation}

\noindent is the Lorentz force per unit volume, $P$ the pressure,
$\Sigma$ the surface mass density, ${\bf v}$ the flow velocity and
${\bf B}$ the magnetic field ($\mu_0$ is the permeability of vacuum).
SI units are used throughout the paper.  

To close the system of equations, we adopt a barotropic equation of
state:

\begin{equation}
P =  P  \left( \Sigma \right).
\label{state}
\end{equation}

\noindent The sound speed $c$ is then given by:

\begin{equation}
c^2 = \frac{d P }{d\Sigma}.
\label{csound}
\end{equation}

\subsection{Equilibrium model}

We adopt a (non--rotating) Cartesian coordinate system $(x,y,z)$ and
denote $(\hat{\bf x}, \hat{\bf y}, \hat{\bf z})$ the associated unit
vectors.  We consider a Cartesian shear flow in the $(x,y)$ plane in
which the velocity ${\bf v}= v(x) \; \hat{\bf y}$ is in the
$y$--direction and has a gradient along $x$.  We assume that the
equilibrium configuration contains a uniform magnetic field ${\bf
B}=B_0 \; \hat{\bf y}$ in the $y$--direction.  There is no Lorentz
force acting on the flow at equilibrium.  We also assume that $\Sigma$
and $P$ are uniform.

\subsection{Response to a small perturbation}

We consider a perturbation with fixed frequency $\omega$ propagating
in the $y$--direction with wavenumber $k_y$.  Because the equilibrium
flow is steady and independent of $y$, we can expand each of the
perturbed quantities in Fourier series with respect to the variables $t$ and 
$y$ and solve separately for each values of $\omega$ and $k_y$.  The 
general problem may then be reduced to calculating the response of the flow 
to the real part of a complex perturbation proportional to $\exp \left[ i \left(
k_y y-\omega t \right) \right]$.

The variables $x$ and $y$ may be seen as the equivalent of the
cylindrical coordinates $r$ and $\phi$, respectively, in a disc with
$r \rightarrow \infty$.  If the perturbation is due to an embedded
planet on a circular orbit in the disc, $\omega=m \Omega_p$, where
$\Omega_p$ is the angular velocity of the planet and $m$ is the
azimuthal mode number (varying from 0 up to infinity), and $k_y=m/r$.
Finite values of $k_y$ correspond to $m \rightarrow \infty$ (as we are 
considering the limit $r \rightarrow \infty$ in this analysis).

We denote Eulerian perturbations with a prime.  We make all Eulerian
fluid state variable perturbations complex by writing:

\begin{equation}
X'(x, y, t) = \sum_{k_y=0}^{\infty} X'_{k_y} (x) e^{i \left(
k_y y - \omega t \right)} ,
\end{equation}

\noindent where $X$ is any state variable.  The physical perturbations
will be recovered by taking the real part of these complex quantities.
We denote $\xib$ the Lagrangian displacement, and write its $k_y$--th
Fourier component as $\xib_{k_y} (x) \exp \left[ i \left( k_y y - \omega
t \right) \right]$.  Since here we are only interested in the
perturbations in the plane of the initial flow, we take $\xi_z=0$.

Linearization of the induction equation~(\ref{induction}) gives $B'_x
= ik_yB_0 \xi_x$ and $B'_y = -B_0 \partial \xi_x / \partial x$, where we have
used $v'_x = i \left( k_y v- \omega \right) \xi_x$.  The linearized
Lorentz force can then be calculated from equation~(\ref{lorentz}):

\begin{equation}
{\bf F}' = \frac{B_0^2}{\mu_0} \left( \frac{\partial^2 \xi_x}{\partial x^2}
-k_y^2 \xi_x \right) \hat{\bf x} .
\end{equation}

We now linearize the equation of motion~(\ref{motion}) and the
equation of continuity~(\ref{mass}).  Using equations~(\ref{state})
and~(\ref{csound}), we get:

\begin{equation}
\frac{\left( v- v_\varphi \right)^2}{c^2} 
k_y^2 \xi_{x,k_y} = \frac{1}{\Sigma}
 \frac{d\Sigma'_{k_y}}{dx} - \frac{F'_{x,k_y}}{\Sigma c^2} ,
\end{equation}

\begin{equation}
\frac{\left( v- v_\varphi \right)}{c} \left( \frac{v'_{y,k_y}}{c} + 
\frac{1}{c} \frac{dv}{dx} 
\xi_{x,k_y} \right) = - \frac{\Sigma'_{k_y}}{\Sigma},
\end{equation}

\begin{equation}
\left( v- v_\varphi \right)  \frac{\Sigma'_{k_y}}{\Sigma} + 
\frac{\partial \left[ \left( v- v_\varphi \right) \xi_{x,k_y}
\right]}{\partial x} +  v'_{y,k_y} = 0.
\end{equation}

\noindent Here $ v_\varphi \equiv \omega /k_y$ is the phase velocity of
the perturbation in the $y$--direction.  We can further eliminate
$v'_{y,k_y}$ and $\Sigma'_{k_y}$ to get the following second--order
differential equation for $\xi_{x,k_y}$:

\begin{equation}
{\cal A}_2 \frac{\partial^2 \xi_{x,k_y}}{\partial x^2} + {\cal A}_1
\frac{\partial \xi_{x,k_y}}{\partial x} + {\cal A}_0 \xi_{x,k_y} = 0 ,
\label{EQ2}
\end{equation}

\noindent with:

\begin{equation}
{\cal A}_2 = 1 + \frac{1}{ \beta} \left[ 1 - \frac{c^2}{ \left( v-
v_\varphi \right)^2} \right],
\label{coefA2}
\end{equation}

\begin{equation}
{\cal A}_1 = \left[ 1 - \frac{\left( v- v_\varphi \right)^2}{c^2} 
\right]^{-1} \frac{2 }{ v- v_\varphi } \frac{dv}{dx} ,
\end{equation}

\begin{equation}
{\cal A}_0 = \left[ \frac{\left( v- v_\varphi \right)^2}{c^2} - 1 
\right] \left[ 1 - \frac{1}{\beta} \frac{c^2}{\left( v- v_\varphi
\right)^2} \right] k_y^2 ,
\end{equation}

\noindent where $\beta \equiv c^2 / v_A^2$, with $v_A \equiv
\sqrt{B_0^2 / \left( \mu_0 \Sigma \right)}$ being the Alfv\'en speed
($B_0^2$ in the expression for $v_A$ should be thought of as the square
of the magnetic field integrated over the thickness of the fluid).
Note that $\beta$ is usually taken to be the ratio of the thermal to
magnetic pressure, which is a factor of two larger than the parameter we
use here.

We are now going to determine the regions of space where waves can
propagate along the $x$--direction, i.e. radially in a disc.  Note
that all the perturbations considered here do propagate along the
$y$--direction (azimuthally in a disc) for all values of $x$.  We
begin with the no shear case to gain some understanding in the
dynamics of the waves.

\subsubsection{Case $dv/dx=0$:}

Here ${\cal A}_1=0$ and $\xi_{x,k_y}$ is a wave propagating in the $x$
direction if $ {\cal A}_2 {\cal A}_0 >0$.  This corresponds to $
\left|v-v_\varphi \right| > \max (c, v_A)$ or $c/\sqrt{\beta+1} <
\left|v-v_\varphi \right| < \min(c, v_A)$.  When this is satisfied,
$\xi_x = \xi_{x,k_y} \exp \left[ i \left( k_y y - \omega t \right) \right]$
is a dispersive wave propagating along an oblique direction.

When $k_y=0$, $\xi_x$ is a longitudinal magneto--acoustic wave (fast
mode) propagating along the $x$--direction with a velocity
$\left( v_A^2+c^2 \right)^{1/2}$.

For $B=0$, i.e. $\beta \rightarrow \infty$, we have $\xi_x \propto
\exp \left[ i \left( k_x x + k_y y - \omega t \right) \right]$, with
$k_x^2 = k_y^2 \left[ \left(v- v_\varphi \right)^2/c^2 -1 \right]$.
The wavenumber $k=\left(k_y^2 + k_x^2 \right)^{1/2}$ then satisfies
$k= k_y \left|v-v_\varphi \right| /c$, which is characteristic of an
oblique sound wave which in general is dispersive (it is
non--dispersive only when $v=0$).


\subsubsection{Case $dv/dx \ne 0$ and $B=0$:}

We will suppose here that $dv/dx \ne 0$ but $d^2v/dx^2 = 0$.  Note
that in a Keplerian disc with $r \rightarrow \infty$, the second
derivative of the velocity is $\sim 1/r$ times the first derivative,
so that it can be neglected (e.g. the shearing--sheet).

We define the new variable:

\begin{equation}
u_{x,k_y} \equiv \xi_{x,k_y} \exp \left( \frac{1}{2} 
\int \frac{{\cal A}_1}{{\cal A}_2} dx \right) .
\end{equation}

Equation~(\ref{EQ2}) can then be written under the form:

\begin{equation}
\frac{\partial^2 u_{x,k_y}}{\partial x^2} +
{\cal K} u_{x,k_y} = 0 ,
\label{EQ2b}
\end{equation}

\noindent with 
\begin{equation}
{\cal K} = \frac{{\cal A}_0}{{\cal A}_2} - \frac{1}{4} \left(
\frac{{\cal A}_1}{{\cal A}_2} \right)^2 - \frac{1}{2} \frac{d}{dx}
\left( \frac{{\cal A}_1}{{\cal A}_2} \right) .
\label{coefK}
\end{equation}

When $B=0$, i.e. $\beta \rightarrow \infty$, this gives:

\begin{equation}
{\cal K} = \left[ 1 - \frac{\left( v- v_\varphi \right)^2}{c^2}
\right]^{-2} k_y^2 
\left\{
-\frac{3}{k_y^2c^2} \left( \frac{dv}{dx} \right)^2 + \left[ \frac{\left(
v- v_\varphi \right)^2}{c^2} -1 \right]^3 \right\} .
\end{equation}

The solution of equation~(\ref{EQ2b}) is a wave propagating in the
$x$--direction if ${\cal K}>0$, i.e.:

\begin{equation}
\frac{ \left( v - v_\varphi \right)^2}{c^2} > \left[ \frac{3}{k_y^2 c^2}
\left( \frac{dv}{dx} \right)^2  \right]^{1/3} + 1.
\label{cond1}
\end{equation}

\begin{figure}
\centerline{
\epsfig{width=10.cm, angle=270, file=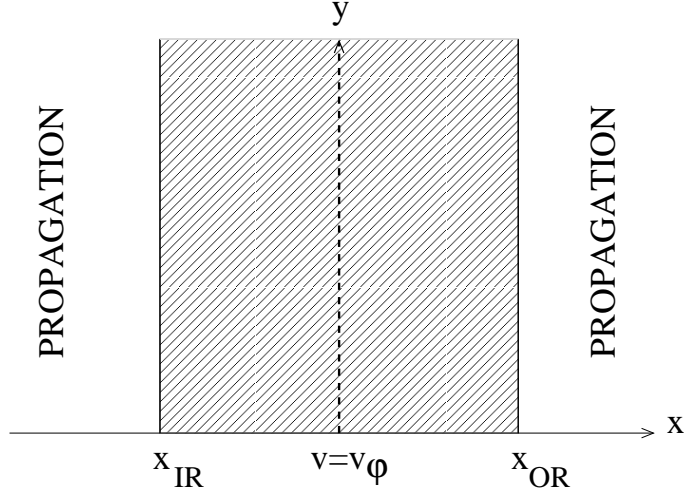} 
}
\caption[]{Wave propagation in a cartesian shear flow.  Case where
$dv/dx \ne 0$ and $B=0$: The shaded area indicate the regions where
the waves are evanescent in the $x$--direction (they propagate only in
the $y$--direction).  $v_{\varphi} \equiv \omega/k_y$ is the phase
velocity along the $y$--direction.  The location $v=v_\varphi$ is not
a singularity. Waves propagate in the $x$--direction beyond the
turning points $x=x_{IR}$ and $x=x_{OR}$ given by
equation~(\ref{cond1}). }
\label{fig1}
\end{figure}

\noindent Note that if $dv/dx=0$, we recover the condition $
\left|v-v_\varphi \right| > c$.  There is no singularity at
$v=v_\varphi$.  Only $\partial \xi_{x,k_y}/\partial x=0$ there.  The
inequality~(\ref{cond1}) is the equivalent in a shear flow of the
criterion for density wave propagation in a Keplerian disc:
$(v-v_p)^2/c^2 > \kappa^2 /(k_y^2 c^2)$ where $v_p$ is the orbital
velocity at corotation and $\kappa$ is the epicyclic frequency
($\kappa^2 = v^2/r^2 + {\rm pressure \; term}$).  In
Figure~\ref{fig1}, the shaded area indicates where the waves are
evanescent in the $x$--direction (the perturbation does propagate
along the $y$--direction for all values of $x$).  We note $x_{IR}$ and
$x_{OR}$ the location of the turning points, beyond which waves
propagate in the $x$--direction ('IR' and 'OR' stand for inner and
outer 'resonances', respectively, by analogy with the Lindblad
resonances in a disc).

In the limit $\left| v-v_\varphi \right| \gg c$, i.e. far enough from
the turning points, the wavenumber in the $x$--direction, $k_x = {\cal
K}^{1/2}$, can be approximated by $k_x \simeq k_y \left| v-v_\varphi
\right| / c \gg k_y$.  This is characteristic of dispersive sound
waves whose group velocity in the $x$--direction is $c$ and whose
wavelength along $x$ decreases as they propagate outward.  Again, this
is equivalent to the density waves that propagate away from the
Lindblad resonances in a Keplerian disc.

\subsubsection{Case $dv/dx \ne 0$ and $B \ne 0$:}
\label{sec:analysis}

Here again we suppose $d^2v/dx^2=0$.  As in the previous subsection,
we have to solve equation~(\ref{EQ2b}) with ${\cal K}$ given by
equation~(\ref{coefK}).  It can be shown that:

\begin{equation}
{\cal K} = \frac{k_y^2}{{\cal A}_2} 
\left[ \frac{\left( v-v_\varphi \right)^2}{c^2} - 1 \right]^{-2} {\cal D} ,
\end{equation}

\noindent where we have defined:

\begin{equation}
{\cal D} \equiv - \frac{3}{k_y^2 c^2}\left( \frac{dv}{dx} \right)^2 +
\left[ \frac{\left( v-v_\varphi \right)^2}{c^2} - 1 \right]^3 \left[ 1 -
\frac{1}{\beta} \frac{c^2}{\left( v-v_\varphi \right)^2} \right].
\end{equation}

\noindent Then the condition ${\cal K}>0$ is equivalent to ${\cal A}_2
{\cal D} >0$.  We are going to study in turn the conditions ${\cal D}
>0$ and ${\cal A}_2 >0$.

We first suppose $\left| v-v_\varphi \right|/c \ll 1$.  Then ${\cal
D}>0$ is equivalent to:

\begin{equation}
\frac{\left( v-v_\varphi \right)^2}{c^2} < \frac{1}{\beta}
\left[ \frac{3}{k_y^2 c^2}\left( \frac{dv}{dx} \right)^2 + 1 \right]^{-1}.
\label{cond2}
\end{equation}

\noindent If $\beta \gg 1$ (weak magnetic field) and/or $\left| dv/dx
\right| \gg k_y c$, this condition is consistent with $\left| v-v_\varphi
\right|/c \ll 1$.  We note $x'_{IR}$ and $x'_{OR}$ the turning points
corresponding to this condition.

\noindent We now suppose $\left| v-v_\varphi \right|/c$ is at least on
the order of unity.  Then, if $\beta \gg 1$ and/or $\left| v-v_\varphi
\right|/c \gg 1$, ${\cal D}>0$ is equivalent to:

\begin{equation}
\frac{\left( v-v_\varphi \right)^2}{c^2} > 
\left[ \frac{3}{k_y^2 c^2}\left( \frac{dv}{dx} \right)^2 \right]^{1/3}
+1.
\label{cond3}
\end{equation}

\noindent Again, this calculation is self-consistent if $\beta \gg 1$
and/or $\left| dv/dx \right| \gg k_y c$.  Note that this condition is the
same as that given by equation~(\ref{cond1}), so that the turning
points are here again $x_{IR}$ and $x_{OR}$.

Finally, ${\cal A}_2>0$ is equivalent to:

\begin{equation}
\frac{\left( v-v_\varphi \right)^2}{c^2} > \frac{1}{\beta +1}.
\label{cond4}
\end{equation}

\noindent Equation~(\ref{EQ2b}) has a (regular) singularity at
$x=x_{IMR}$ and $x=x_{OMR}$ (where 'IMR' and 'OMR' stand for inner and
outer magnetic resonances, respectively, by analogy with the magnetic
resonances in a disc), where ${\cal A}_2=0$, i.e. $\left( v-v_\varphi
\right)^2=c^2/\left( \beta +1 \right)$.  The flow however is well
behaved at these locations, in contrast to the flow in a Keplerian
disc (see T03).  Note that in a Keplerian disc, the position of the
magnetic resonances is given by exactly the same equation, provided
$v_\varphi$ is replaced by the orbital velocity of the planet.

\begin{figure}
\centerline{
\epsfig{width=10.cm, angle=270, file=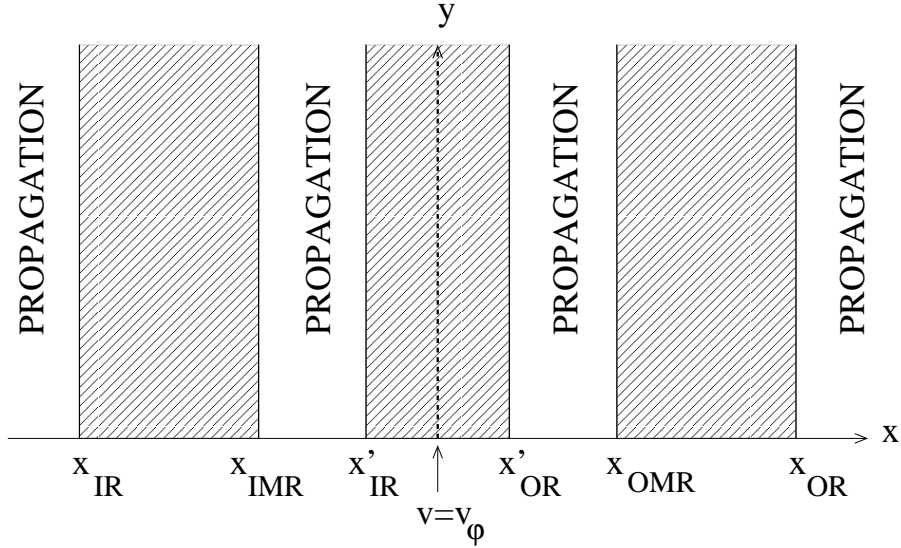} 
}
\caption[]{Same as figure~\ref{fig1} but for $B \ne 0$, and for the
case $\beta \gg 1$ and/or $\left| dv/dx \right| \gg k_y c$.  The turning
points $(x'_{IR}, x'_{OR})$, $(x_{IR}, x_{OR})$ and $(x_{IMR},
x_{OMR})$ are given by equations~(\ref{cond2}), (\ref{cond3}) (or
equivalently eq.~[\ref{cond1}]) and (\ref{cond4}), respectively. Waves
propagate in the $x$--direction beyond the turning points $x_{IR}$ and
$x_{OR}$ but also in a restricted region around the magnetic
resonances $x_{IMR}$ and $x_{OMR}$.  The case represented here
corresponds to $3 \beta \left( dv/dx \right)^2 > k_y^2 c^2$.  When this
inequality is not satisfied, the points $(x'_{IR},x'_{OR})$ and
$(x_{IMR}, x_{OMR})$ have to be swapped.}
\label{fig2}
\end{figure}

In Figure~\ref{fig2}, the shaded areas indicate where the waves are
evanescent in the $x$--direction (the perturbation does propagate
along the $y$--direction for all values of $x$), i.e. where ${\cal
A}_2 {\cal D} <0$.  Waves propagate beyond the turning points that are
also present in a non--magnetized disc, but they propagate as well in
some restricted region around the magnetic resonances.  

In the limit $\left| v-v_\varphi \right| \gg c$, i.e. far enough from
the turning points $x_{IR}$ and $x_{OR}$, the wavenumber in the
$x$--direction, $k_x = {\cal K}^{1/2}$, can be approximated by: 

\begin{equation}
k_x \simeq k_y \left( \frac{\beta}{\beta+1} \right)^{1/2}
\frac{ \left| v-v_\varphi \right|}{c} .
\end{equation}

\noindent Note that, like in the non magnetic case, $k_x \gg k_y$ and
the wavelength in the $x$--direction decreases as the waves propagate
outward.  This dispersion relation is characteristic of
magneto--acoustic waves (fast mode) whose group velocity in the
$x$--direction is $(c^2+v_A^2)^{1/2}$.  Again, this is equivalent to
the waves that propagate away from the Lindblad resonances in a
Keplerian disc (see eq.~[37] and~[38] in T03).

In the vicinity of the magnetic resonances, ${\cal A}_2 \simeq 0$ and
we can write ${\cal K} \simeq {\rm Const}/{\cal A}_2$.  In the regions
where ${\cal K}>0$, the wavenumber in the $x$--direction is $k_x =
{\cal K}^{1/2}$, so that the group velocity in this direction,
$d\omega/dk_x$, can be calculated by using $2k_k dk_x/d\omega \simeq -
({\rm Const}/{\cal A}_2^2) d{\cal A}_2/d\omega$.  This gives:

\begin{equation}
\frac{d\omega}{dk_x} \simeq {\cal A}_2^{3/2} 
\left[ -\frac{3}{k_y c^2} \left( \frac{dv}{dx} \right)^2 
+\frac{v_A}{c} \left( \frac{c^2}{c^2 + v_A^2} \right)^3 \right]^{-1/2}
\left( \frac{c^2}{c^2 + v_A^2} \right)^{5/2} v_A .
\end{equation}

\noindent This is characteristic of a slow MHD wave.  Note that, as
$d\omega/dk_x$ tends to zero as we approach the magnetic resonances,
the waves propagate essentially along the magnetic field line in the
vicinity of these resonances.

To summarize, fast magneto--acoustic waves propagate away from the Lindblad 
resonances while slow MHD wave propagates in the vicinity of the magnetic 
resonances. This situation is very similar to that in a Keplerian disc (T03),
and these results are going to be used to interpret the numerical
simulations that are presented below.

\section{Numerical simulations}

We perform numerical simulations of a low mass protoplanet embedded in
a disc containing a toroidal magnetic field in order to study the
properties of the magnetic resonances and in particular their effect
on the migration of the planet.

\subsection{Numerical setup}

\subsubsection{The method}

We solved the MHD equations given in section~{\ref{mhd equations}}
using a 2D adaptation of the 3D Eulerian code GLOBAL (Hawley \& Stone
1995) and a modified version of NIRVANA (Ziegler \& Yorke 1997). The
algorithms used by both codes are very similar: they solve the above
equations in cylindrical polar coordinates ($r$, $\phi$) using
time-explicit finite differences. The magnetic field is evolved using
the combined Method Of Characteristics and Constrained Transport
(MOC-CT) algorithm which both preserves the divergence of the magnetic
field and accurately describes the propagation of Alfv\'en waves
(Hawley \& Stone 1995).

Because the magnetic resonances lie very close to the orbit of the
planet, a very high resolution is needed to resolve them. According to
the linear theory, the distance between the planet and the magnetic
resonances is (T03):

\beq
\mid r_M-r_{pl} \mid=\frac{2}{3} \frac{H}{\sqrt{1+\beta}} \, ,
\eeq

\noindent
where $r_M$ and $r_{pl}$ are the radii of the magnetic resonances and
of the planet, respectively, and $H$ is the disc semi--thickness. With
the typical values of $H/r_{pl}=0.1$ and $\beta=2$ that we will use
throughout this paper, we get $\mid r_M-r_{pl} \mid / r_{pl} = 3.8
\times 10^{-2}$. In order to accurately describe the magnetic
resonances, we need to have roughly ten grid cells within this
interval. On a uniform grid, when $r_{pl}=1$, this resolution would
require about $550$ grid cells in the radial direction for a typical
disc model whose radii range from $0.4$ to $2.5$. For such a
resolution, each run would take months on a standard desktop
computer. We have followed two routes to solve this problem. On the
one hand, using GLOBAL, we built a non--uniform grid whose resolution
increases in the neighborhood of the planet (see Appendix A for a
detailed description of the grid setup). On the other hand, we ran
NIRVANA on a multi-processor facility using the MPI library.

Using the first approach, we can run high resolution simulations with
modest computational resources. However, the orbital radius of the
planet has to be fixed so that it always remains in the high
resolution part of the grid during the simulations. Note also that the
calculation has to be made in the frame rotating with the planet, in
which case the inertial forces are treated as described by Kley
(1998). In the second case, we require access to powerful computing
facilities, but the planet can move with respect to the grid, and we
can study its actual migration in the disc. With the combination of
both techniques, we have been able to analyze the effect of the
magnetic resonances in detail.

\subsubsection{The disc model}
\label{discmodel}

The disc parameters we used are described in detail in Nelson et
al.~(2000).  We use dimensionless units for our numerical simulations.
The unit of mass is taken to be the mass of the central star,
$M_\star$=1, and the unit of length is taken to be the initial orbital
radius of the protoplanet, $r_{pl}=1$.  We set the gravitational
constant $G=1$.  We will consider a protoplanet with a mass $M_{pl}
\ll M_\star$, so that, in a non self--gravitating disc, the unit of
time can be approximated by $\left[ r_{pl}^3/(GM_\star)
\right]^{1/2}$.  The surface density is constant with radius, with
$\Sigma=3 \times 10^{-3}$. The inner boundary of the disc is located
at $R_{in}=0.4$ and the outer boundary is at $R_{out}=2.5$. This
results in quite a massive disc, and was chosen to give correspondingly
rapid migration rates because of the long run times required by our
high resolution simulations.  In this paper, we only investigated the
case $M_{pl}=1.5 \times 10^{-5}$, which corresponds to $5$~Earth
masses if the mass of the central star is that of the Sun.

The planet orbiting in the disc is modelled as a softened point
mass. The total gravitational potential $\Phi$ that is exerted at any
point of the disc is the sum of the potential of the central point
mass and the potential of the planet:

\begin{eqnarray}
\Phi(r,\phi) = - \frac{GM_\star}{r} - \frac{GM_{pl}}{\left[
r^2+r_{pl}^2-2rr_{pl} \cos(\phi-\phi_{pl})+\epsilon ^2 \right]^{1/2}}
+ \frac{GM_{pl}}{r_{pl}^3} \; \bb{r} \bcdot \bb{r_{pl}} + \Phi_{ind} +
\Phi_{disc} \, .
\label{potential}
\end{eqnarray}

\noindent
Here the planet is located at the point $(r_{pl},\phi_{pl})$.  The
parameter $\epsilon$ is the smoothing length of the gravitational
potential. We took $ \epsilon=0.1H$, so that it is smaller than the
distance between the planet and the magnetic resonances. The third
term in equation~(\ref{potential}) is the indirect potential due to
the planet, and accounts for the fact that the reference frame
centered on the central star is non--inertial. The fourth term,
$\Phi_{ind}$, represents the indirect potential due to the disc (see
Nelson et al. 2000), and is non--zero only in simulations where the
planet is allowed to migrate. The last term, $\Phi_{disc}$,
corresponds to the potential due to the disc self--gravity, and is
also non--zero only for simulations in which the planet migrates. In
this work, we include only the axisymmetric component of the disc
self--gravity, as we are concerned with accurately modelling the
angular velocity of the disc gas and the embedded planet. Thus:

\begin{equation}
\Phi_{disc}(r) = - G \int_{R_{in}}^{R_{out}} \frac{{\overline \Sigma}(r')
\, r' \, dr' \, d \phi' }{\sqrt{r^2 + {r'}^2 - 2 r r' \cos{(\phi-\phi')} + 
\epsilon^2}}
\label{selfgrav}
\end{equation}

\noindent where ${\overline
\Sigma}(r)=\int \Sigma(r,\phi) d\phi / (2 \pi)$. In this paper, we
consider quite a massive disc model, as it produces more rapid
migration rates. This means, however, that the disc gravity then makes
a non--negligible contribution to the angular velocity of the disc
material and embedded planet. The importance and effects of including
$\Phi_{disc}$ are discussed further in section~\ref{Migration}.

The equation of state is locally isothermal. The vertically integrated
pressure $P=\Sigma c^2$ is calculated with the thin--disc approximation:

\begin{equation}
P =  \left( \frac{H}{r} \right)^2  \frac{GM_\star}{r} \; \Sigma  \, .
\end{equation}

\noindent
The magnetic field is initially purely toroidal. Its strength is
defined through the parameter $\beta$, which is the ratio of the
square of the sound speed to the square of the Alfv\'en velocity at
the initial location of the planet. A power--law variation of
$B_{\phi}$ with radius is allowed:

\begin{equation}
B_{\phi}= \sqrt{\frac{\mu_0 P(r=r_{pl})}{\beta}}
\left(\frac{r}{r_{pl}}\right)^{-p} \, .
\label{bfield law}
\end{equation}

\subsubsection{Boundary conditions}

When we tried to use standard radial boundary conditions (outflow or
reflecting, for example), we found that high frequency oscillations in
the magnetic variables quickly started to grow. Eventually, in all of
the cases we tested, these oscillations perturbed the entire disc to
such an extent that no migration signal could be extracted.  To
overcome this problem we developed a damping procedure to reduce the
amplitude of the waves generated by the planet as they approach the
boundary.

Let $f$ be either components of the fluid velocity, and $f_0$ its
equilibrium value. At the end of each timestep, we calculate a new
value for $f$ according to the formula:

\begin{equation}
f= \left\{ \begin{array}{ll}
f_0+(f-f_0) \times 
\exp\left[ -\left(\frac{r-r_i}{\Delta_i}\right)^2 \right]
& \textrm{if $r<r_i$}\\
f &\textrm{if $r \in [r_i,r_o]$} \\
f_0+(f-f_0) \times 
\exp \left[ -\left(\frac{r-r_o}{\Delta_o}\right)^2 \right] 
& \textrm{if $r>r_o$}
\end{array} \right.
\label{damping}
\end{equation}

\noindent
We found that the best results were obtained when $r_i=0.55$ and 
$\Delta_i=0.35$, and $r_o=2$ and $\Delta_o=2$. It is crucial not to apply 
equation~(\ref{damping}) to the magnetic field, because that would result 
in a violation of the constraint $\nab \bcdot \bf{B}=0$. Note that in the 
control hydrodynamic case that we did, this wave damping procedure produced 
almost the same torque on the planet as the standard reflective boundary 
conditions.

In cases when we used the logarithmic grid, we also found that more
dissipation than just the standard artificial viscosity was required
to reduce the level of high frequency noise. This dissipation was
simply modelled by a small kinematic viscosity $\nu$ (see for example
Nelson et al. 2000), which we took equal to $7.5 \times 10^{-6}$. In
our units, this is equivalent to adding an $\alpha$--viscosity
(Shakura \& Sunyaev 1973) with $\alpha=7.5 \times 10^{-4}$. Over the
time of our simulations, this small viscosity does not produce any
significant change on the background disc structure. Its most
important effect, as expected, is to broaden the magnetic resonances
(see below). This viscosity was used in both the GLOBAL and NIRVANA runs.

\subsubsection{Simulations properties}

\begin{table}
\begin{center}
\begin{tabular}{@{}lcccccc}
\hline\hline
Model & $\beta$ & $p$ & resolution & grid type & migration \\
\hline\hline
G1 & $\infty$ & - & $242 \times 258$ & VARIABLE  & No \\
G2 & 2 & 0 & $242 \times 258$ & VARIABLE & No \\
G3 & 2 & 0 & $242 \times 354$ & VARIABLE & No \\
G4 & 2 & 0 & $286 \times 400$ & VARIABLE & No \\
G5 & 2 & 1 & $242 \times 258$ & VARIABLE & No \\
G6 & 2 & 2 & $242 \times 258$ & VARIABLE & No \\
\\
N1 & 2 & 0 & $1000 \times 1000$ & UNIFORM & No \\
N2 & 2 & 0 & $1000 \times 1000$ & UNIFORM & Yes \\
N3 & 2 & 1 & $1000 \times 1000$ & UNIFORM & Yes \\
N4 & 2 & 2 & $1000 \times 1000$ & UNIFORM & Yes \\
\end{tabular}
\end{center}
\caption{Model parameters.  Column $2$ gives $\beta = c^2/v_A^2$ at
the initial location of the planet. Column $3$ gives $p$, the
power-law exponent describing the variations of the magnetic
field. Finally, columns $4$ and $5$ respectively give the resolution
and type of the grid we used (uniform vs variable), while the last
column indicates whether the planet is allowed to migrate or not.}
\label{models}
\end{table}

The parameters of our runs are described in table~\ref{models}. The
first column gives their label. Its first letter indicates the code
that has been used. ``G'' stands for models that have been run with
GLOBAL, while models whose label starts with ``N'' have been performed
with NIRVANA.  Column~2 gives the ratio $\beta = c^2 / v_A^2$ at the
initial orbital radius of the protoplanet and column~3 gives the
exponent $p$ of the magnetic field power law (see eq.~[\ref{bfield
law}]). Columns~4 and 5 describe the grid properties, giving the
resolution and the type of the grid (uniform vs logarithmic). Finally,
the last column states whether the planet is allowed to migrate
through the disc or stay on a fixed orbit during its evolution. The
initial coordinates of the planet are $(r_{pl},\phi_{pl})=(1,\pi)$. In
the remaining of the paper, all times are measured in units of its
initial orbital period $P=2\pi$.

\subsection{Planet fixed on a circular orbit}

In the simulations presented in this section, the planet is kept on a
fixed circular orbit.  The torque it exerts on the disc is computed,
but the feedback torque from the disc on the planet is ignored.

\subsubsection{Case of a uniform field}

\begin{figure}
\centerline{
\epsfig{width=7.cm, angle=0, file=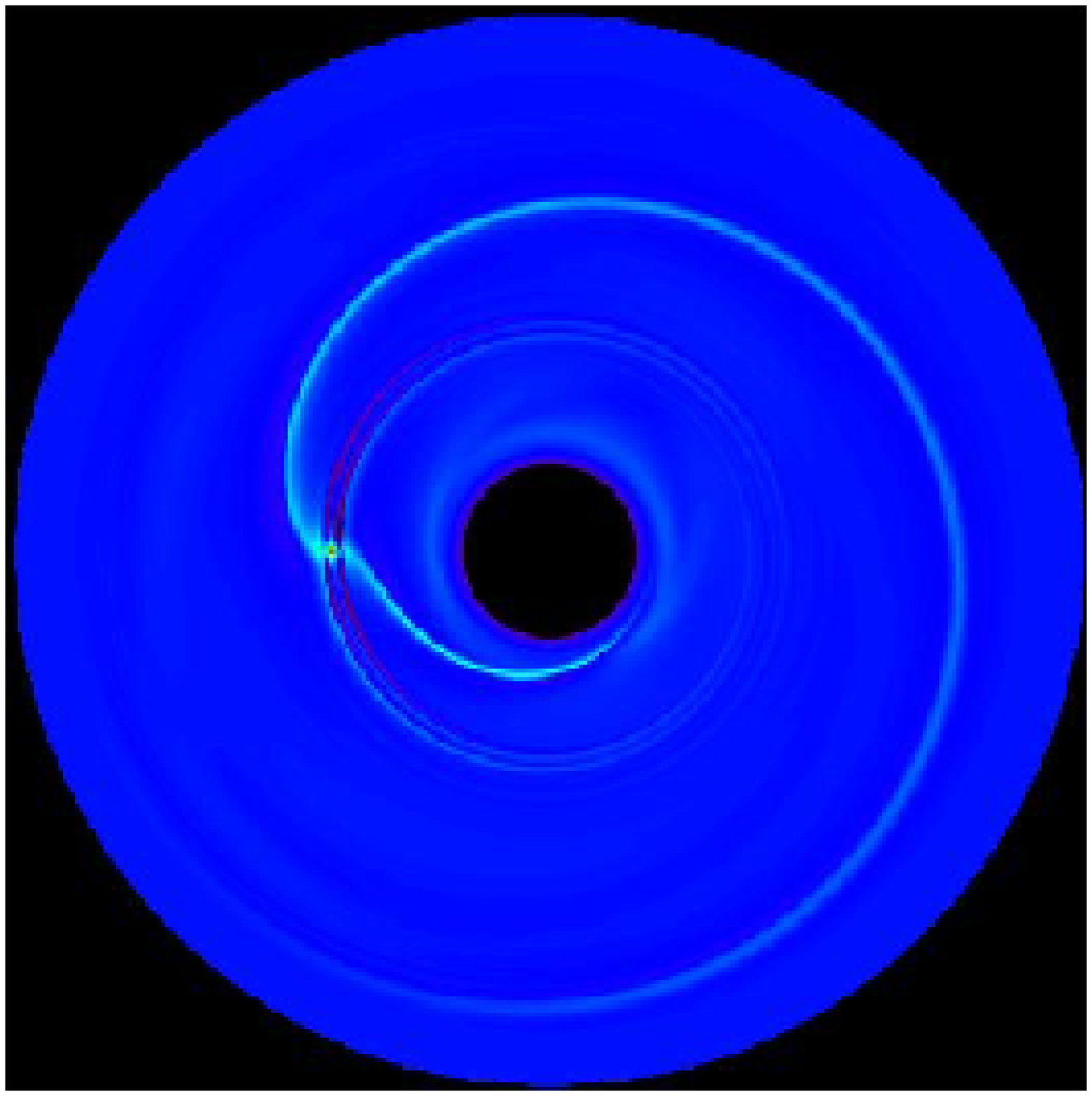} 
\epsfig{width=7.cm, angle=0, file=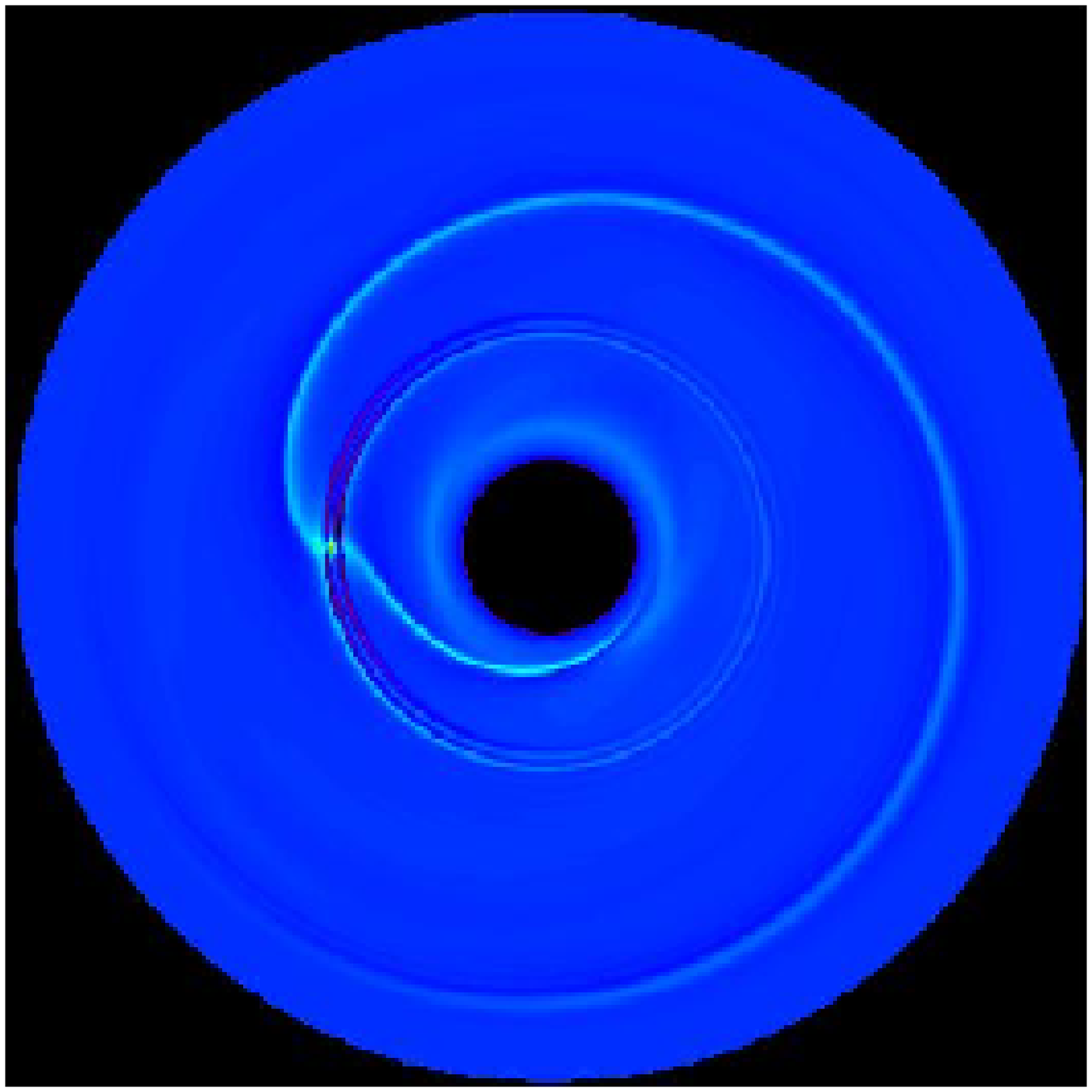} 
}
\caption[]{Disc surface density at time $t=5$ ({\it left panel}) and 
$t=10$ ({\it right panel}) in model G2. The slow MHD waves are seen 
propagating along the magnetic field lines, at the radius of the magnetic 
resonances.}
\label{fig3}
\end{figure}

\noindent
We first begin with a description of our standard case, model G2. It
starts with a uniform magnetic field in the disc, such that $\beta=2$
at the location of the planet. Using the logarithmic grid, the
resolution in the neighborhood of the planet is $\Delta r=3 \times
10^{-3}$, which gives about $12$ cells between the orbit of the planet
and the predicted position of the magnetic resonances. We ran this
model for $160$ orbits at the location of the planet.

When the simulation starts, the gravitational potential of the planet
induces a perturbation on the disc. Fast magnetosonic waves, namely
density waves modified by the magnetic pressure, are launched at the
Lindblad resonances (defined as $x_{IR}$ and $x_{OR}$ in
section~\ref{sec:analysis}) and propagate radially away from the
planet orbit. Their behavior is very similar to that of density waves
in the hydrodynamic case. Slow MHD waves also appear. They propagate
azimuthally, along the magnetic field line, mostly at the expected
location of the magnetic resonances (defined as $x_{IMR}$ and
$x_{OMR}$ in section~\ref{sec:analysis}). This is illustrated by
figure~\ref{fig3}, which shows the surface density in the disc at
times $t=5$ ({\it left panel}) and $t=10$ ({\it right panel}). The
slow MHD waves are seen propagating around the disc. They are very
tightly wrapped, indicating that they undergo very little radial
propagation.

This is in complete agreement with the analysis presented in
section~\ref{sec:analysis} and the results displayed in
figure~\ref{fig2}.  We have checked that in the simulations the waves
that propagate around the magnetic resonances are dominated by values
of $m$ in the range 0--5.  In that case, the approximation $|dv/dx| \gg
k_y c$ used in the analysis of section~\ref{sec:analysis} is
satisfied, and we are in the case displayed in figure~\ref{fig2} where
the points $(x'_{IR},x'_{OR})$ are inside the magnetic resonances.

\begin{figure}
\centerline{
\epsfig{width=8.cm, angle=0, file=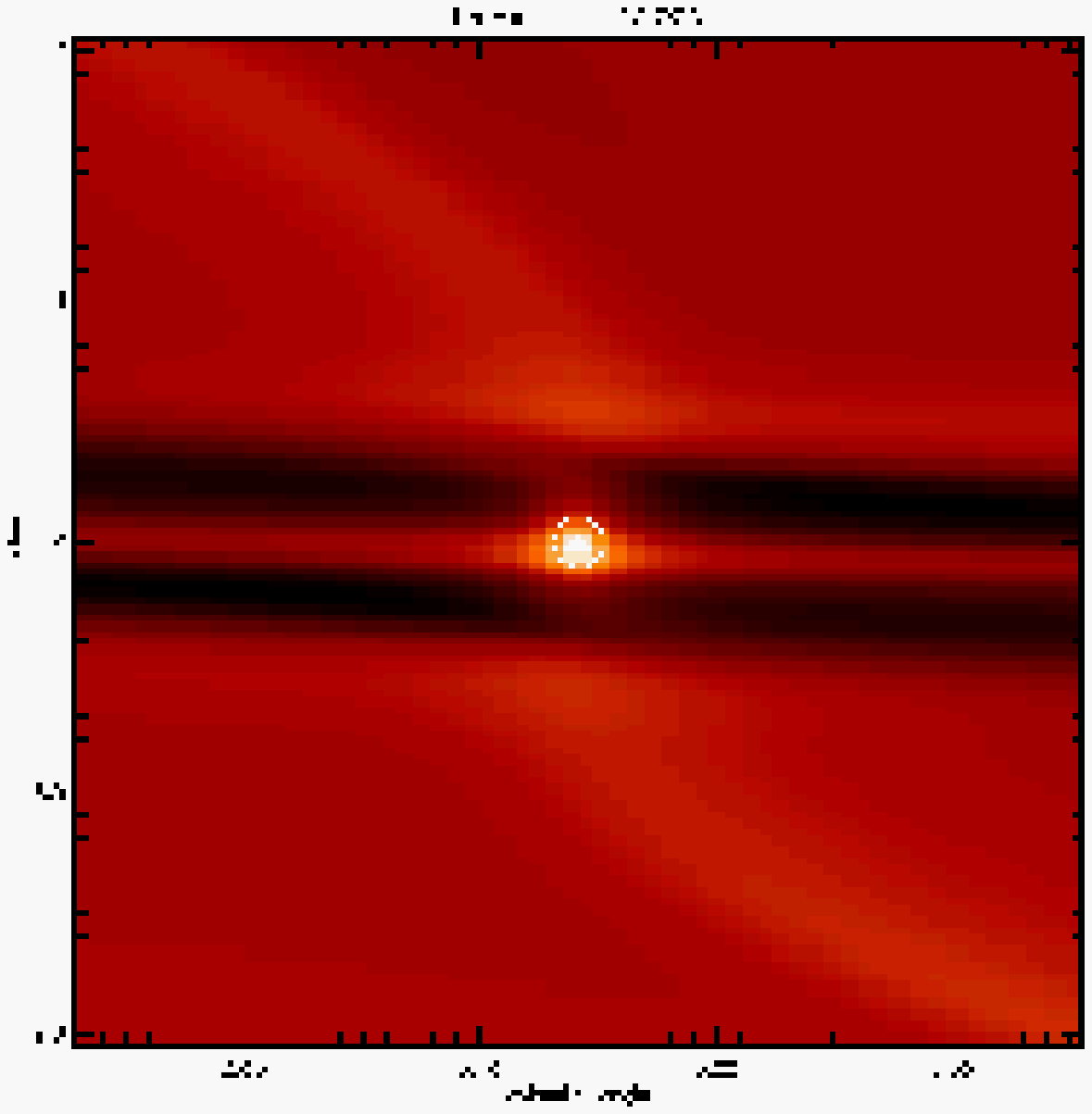} 
\epsfig{width=8.cm, angle=0, file=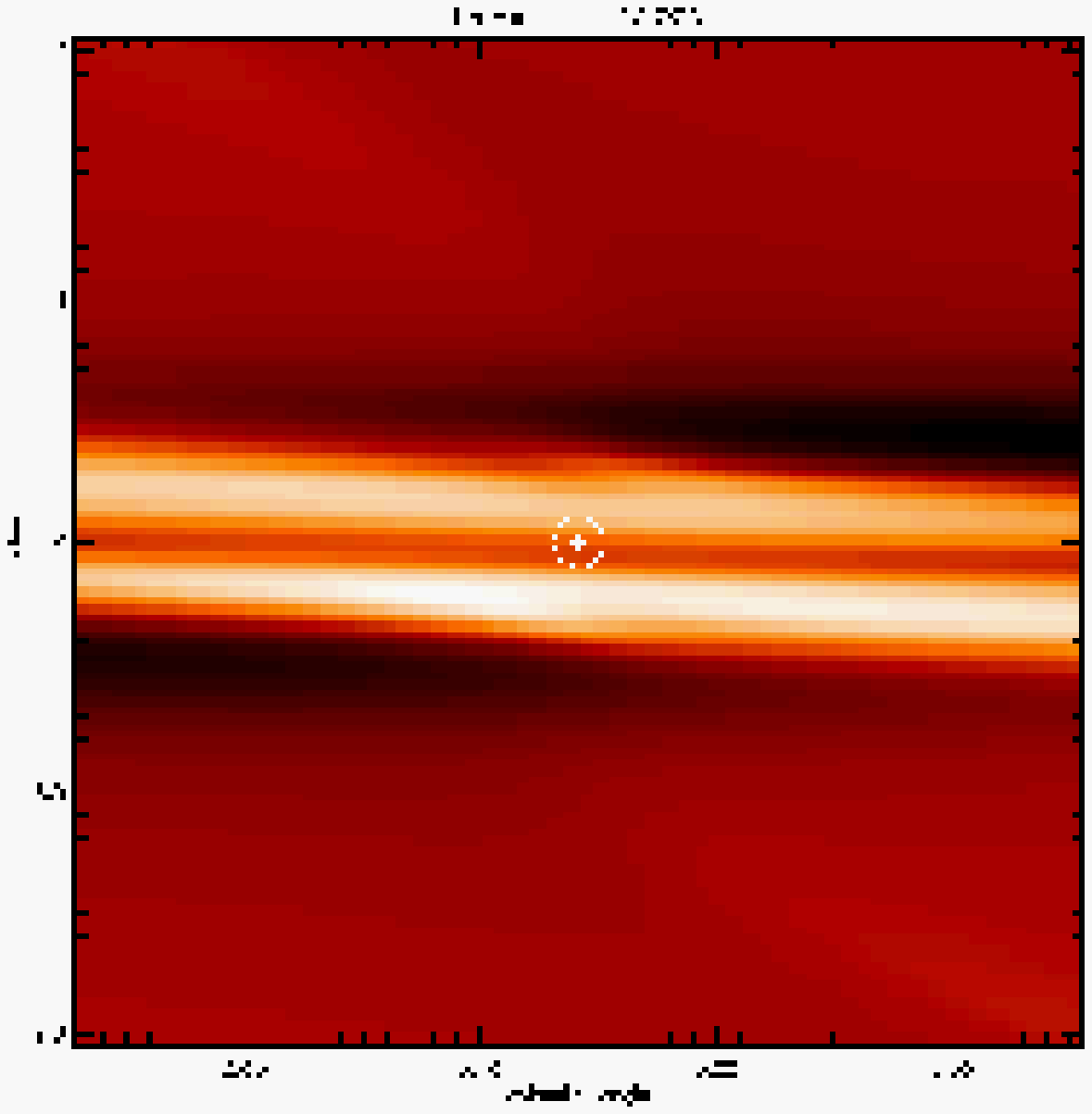} 
}
\caption[]{Snapshot of the density ({\it left panel}) and the toroidal 
component of the magnetic field $B_{\phi}$ ({\it right panel}) for model G2 
after $120$ orbits. The white cross shows the position of the planet, while 
the radius of the circle is equal to the smoothing length. The magnetic 
resonances appear on both sides of the planet's orbit as a decrease in the 
density and an increase in $B_{\phi}$.}
\label{fig4}
\end{figure}

The magnetic resonances appear very clearly on both sides of the
planet's orbit.  This is shown in figure~\ref{fig4}. The left panel
represents the surface density in the vicinity of the planet at time
$t=120$, while the right panel shows $B_{\phi}$. The 
magnetic resonances are apparent as a decrease in the density and an
increase in the magnetic field. Note that this calculation is
performed close to the linear regime. The perturbation of the
variables is small compared to the equilibrium values. Indeed, the
increase of the magnetic field strength is $18$\% and the decrease of
the density is only $4$\%. This allows a close comparison between
these simulations and the analytical results obtained in the linear
regime (see section~\ref{sec:analysis} and below).  As noted above,
the location of the magnetic resonances for $\beta=2$ is $r_M$ such
that $|r_M - r_{pl}|/r_{pl} = 3.8 \times 10^{-2}$.  This agrees with
figure~\ref{fig4}, from which we can also see that the perturbation is
much more important inside these resonances than outside.  This
suggests that the turning points $x'_{IR}$ and $x'_{OR}$ defined in
section~\ref{sec:analysis} are inside the resonances, i.e. waves
propagate in a restricted region inside the resonances, as illustrated
in figure~\ref{fig2}.

An important goal of these simulations is to investigate the effect of
the magnetic resonances on the torque exerted by the disc on the
planet.  The torque ${\bf T}_{pl}$ exterted by the planet on the disc
is:

\begin{equation}
{\bf T}_{pl}= - \int \!\!\! \int_{disc} \Sigma {\bf r} \tim \frac{G
M_{pl} \left( {\bf r}-{\bf r_{pl}} \right)} {\mid {\bf r}-{\bf r_{pl}}
\mid^{3}} \, r \, \textrm{d}r \, \textrm{d}\phi
\end{equation} 

\noindent The torque ${\bf T}$ exerted by the disc on the planet is
$-{\bf T}_{pl}$.  Its value, normalized by the quantity $GM_{pl}
r_{pl}$, is:

\beq T= \int \!\!\! \int_{disc} \frac{\Sigma r^2 \sin(\phi-\phi_{pl})
\, \textrm{d}r \, \textrm{d}\phi}{\mid {\bf r}-{\bf r_{pl}} \mid^{3}} \; .
\eeq

\begin{figure}
\centerline{
\epsfig{width=12.cm, angle=0, file=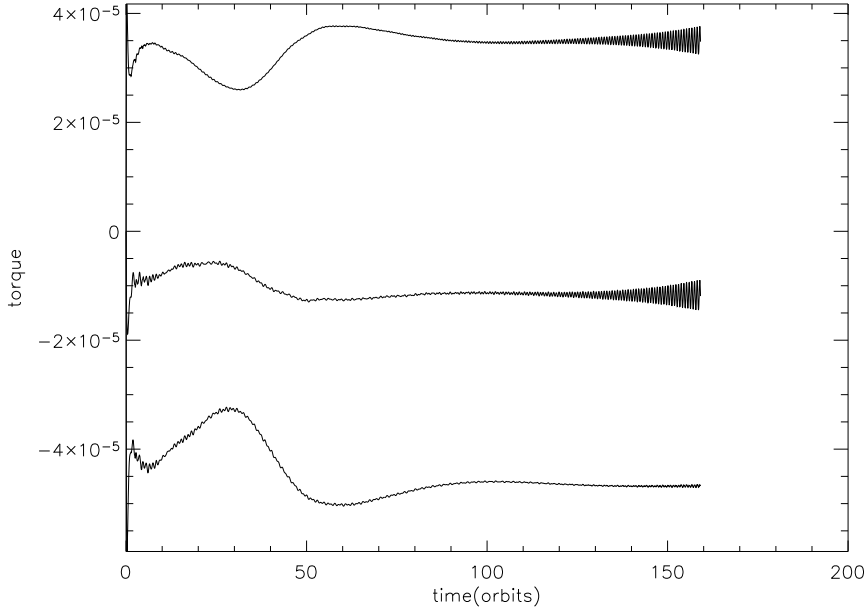} 
}
\caption[]{Time history of the torque exerted by the disc on the planet 
for model G2. From bottom to top, the curves respectively show 
the outer torque, the total torque and the inner torque. Since the total 
torque is negative, the planet migrates inward.}
\label{fig5}
\end{figure}

\noindent 
Figure~\ref{fig5} shows the time history of $T$ for model G2: the
upper curve plots the torque exerted on the planet by the inner part
of the disc ($r<r_{pl}$). It is positive, which indicates that the
inner disc tends to produce outward migration. The lower curve shows
the torque exerted by the outer disc ($r>r_{pl}$) on the planet. The
middle curve is the sum of the two, i.e. the total torque exerted by
the disc on the planet. One can see that it saturates at a constant
and negative value after $80$ orbits. Note also the high frequency
variations appearing at the end of the simulation (for $t>140$). They
are due to our particular treatment of the inner boundary
condition. However, they do not influence the overall behavior of the
disc. As in the hydrodynamic case, figure~\ref{fig5} demonstrates
that the disc causes the planet to migrate inward. This is because the
inner and outer magnetic resonances balance each other in that case.

\begin{figure}
\centerline{
\epsfig{width=12.cm, angle=0, file=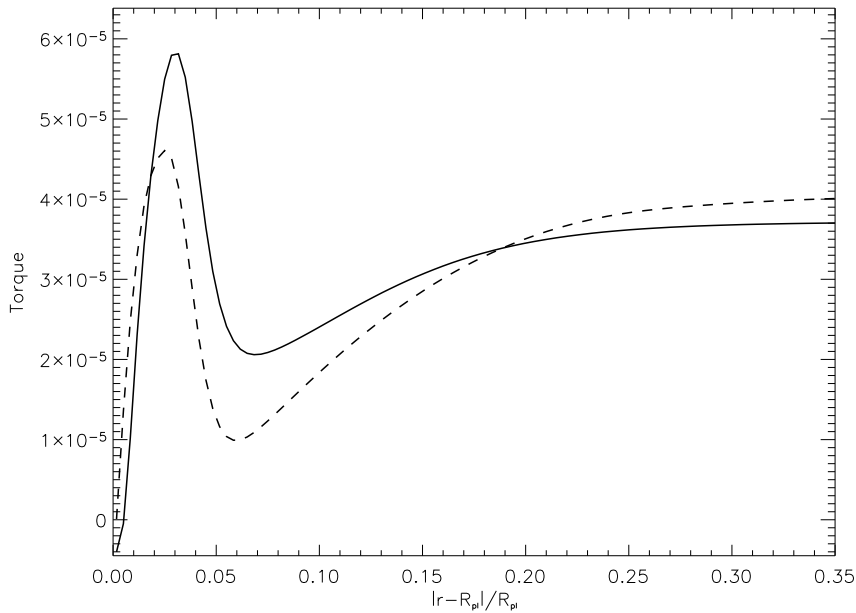} 
}
\caption[]{Torque exerted on the planet by the annulus with radii
between $r_{pl}$ and $r$ as a function of $|r-r_{pl}|/r_{pl}$ for
model G2.  The solid line shows $T(r<r_{pl})$ and the dashed line
$-T(r>r_{pl})$.  The magnetic resonances are obvious at
$|r-r_{pl}|/r_{pl} \simeq 0.03$. The expected value from linear theory
is $0.038$.}
\label{fig6}
\end{figure}

Nevertheless, both resonances still affect the torque, as is
illustrated in figure~\ref{fig6}. The dashed line shows the quantity
$-T(r>r_{pl})$, which is the opposite of the torque exerted on the
planet by an annulus with radii between $r_{pl}$ and $r>r_{pl}$ as a
function of $|r-r_{pl}|/r_{pl}$ for model G2. The solid line shows the
inner torque $T(r<r_{pl})$. On both curves, the magnetic resonances
show up as a peak in these quantities. Their radii is around 0.03.
This is in reasonable agreement with linear theory, which gives $\mid
r-r_M \mid =0.038$ for the parameters of model G2. Note that their
width is larger than expected in an invicid disc because of the
$\alpha$--type viscosity used here.  This is probably why they are not
exaclty at 0.038.  Runs performed in an inviscid disc for a shorter period 
indeed show the resonances at their expected radii (Fromang 2004).

\begin{figure}
\centerline{
\epsfig{width=12.cm, angle=0, file=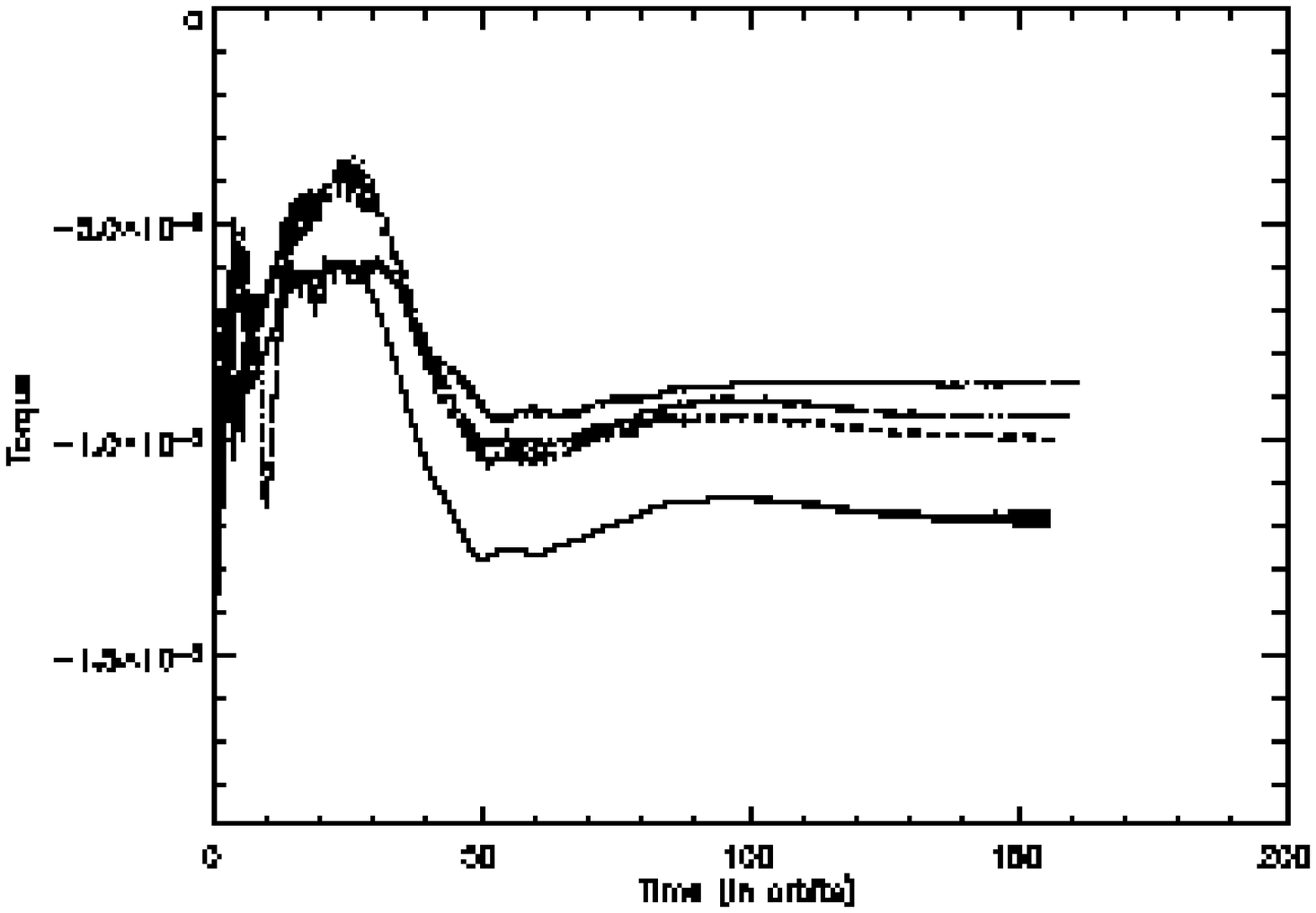} 
}
\caption[]{Comparison of the total torque exerted by the disc on the planet 
for model G2 ({\it solid line}), G3 ({\it dashed line}), G4 
({\it dotted line}) and N1 ({\it dotted-dashed line}). In all cases, the 
saturated torques lie between $-8.5 \times 10^{-6}$ and $-1.2 \times 10^{-5}$. 
This indicates that the results presented in this paper only weakly depend on 
the code and on the resolution.}
\label{fig7}
\end{figure}

In figure~\ref{fig7} we plot the torque corresponding to a variety of
simulations performed with differing resolutions. Results from runs
G2, G3, and G4 are plotted here, showing modest dependence of the
torque on resolution.

Also plotted in figure~\ref{fig7} are the torques calculated from run
N1 performed using NIRVANA. Here a uniform grid in $r$ and $\phi$ was
used, with ($N_r$, $N_{\phi}$)=(1000, 1000) grid cells being employed.
This gives a resolution of $\Delta r = 2.1 \times 10^{-3}$, so that
there are $\sim 18$ grid cells between the location of the planet and
the predicted location of the magnetic resonances. The model was run
with the planet being held on a fixed circular orbit, in a rotating
frame that corotates with the planet, for a time of 160 orbits.  The
resulting torques are in very good agreement with those obtained
during runs G2 , G3 and G4.

\subsubsection{Case of a non--uniform field}
\label{nonuniform simus}

In this section, we investigate the effect of a radial gradient of the
magnetic field on the migration properties of the protoplanet. To do
so, we ran $3$~MHD simulations, models G2 (described above), G5 and
G6, corresponding to $p=0,1,2$ respectively (see eq.~[\ref{bfield
law}]).  All of the other parameters of the three simulations are the
same. We also compare these models with model G1, a case without a
magnetic field ($B_{\phi}=0$).

\begin{figure}
\centerline{
\epsfig{width=12.cm, angle=0, file=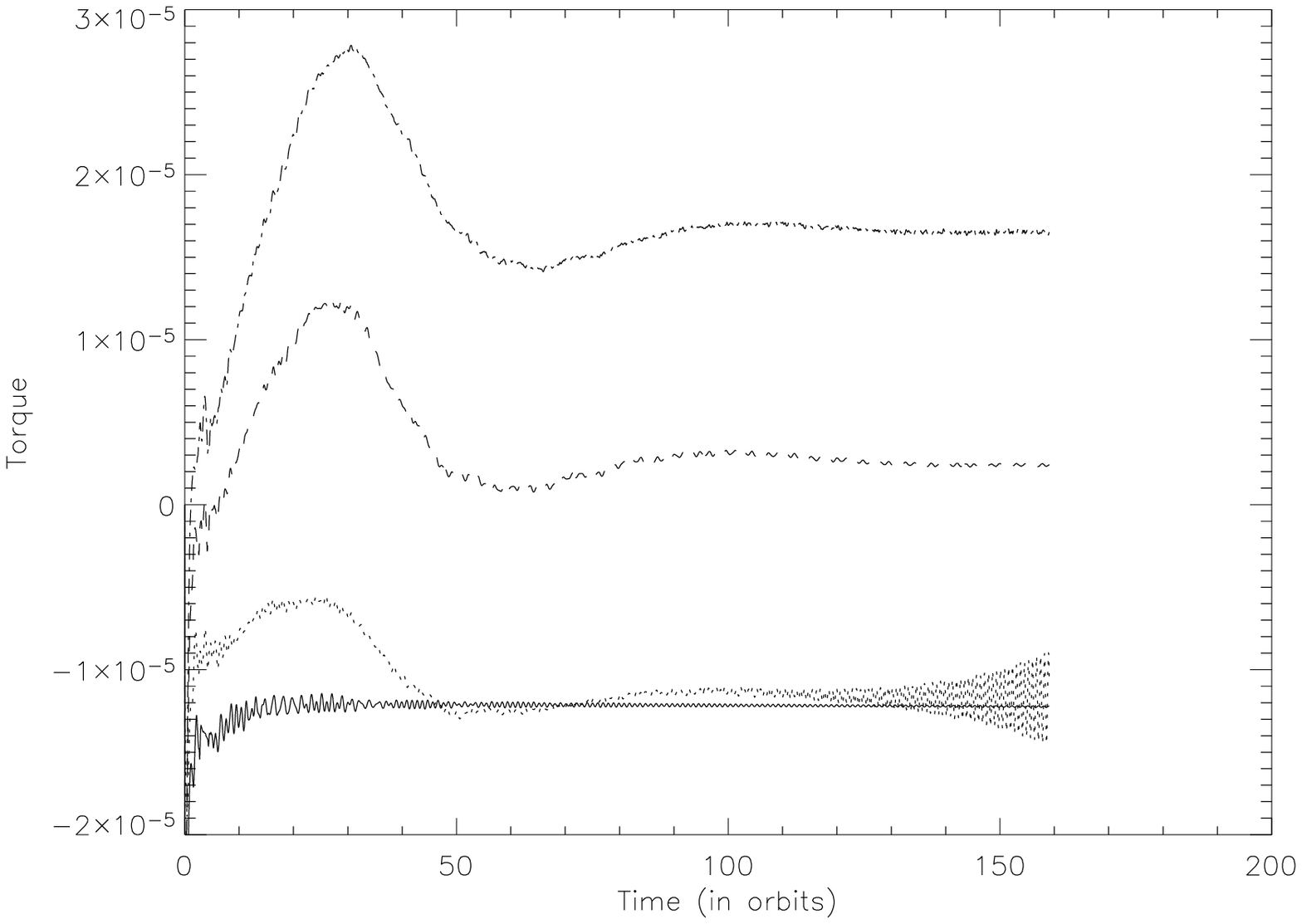} 
}
\caption[]{Total torque $T$ exerted by the disc on the 
planet for model G1 ({\it solid line}), G2 ({\it dotted line}), G5 
({\it dashed line}) and G6 ({\it dotted--dashed line}). $T$ increases 
as the magnetic field steepens and becomes positive when 
$B_{\phi} \propto r^{-2}$: the planet will migrate outward in that case.}
\label{fig8}
\end{figure}

\begin{figure}
\centerline{
\epsfig{width=12.cm, angle=0, file=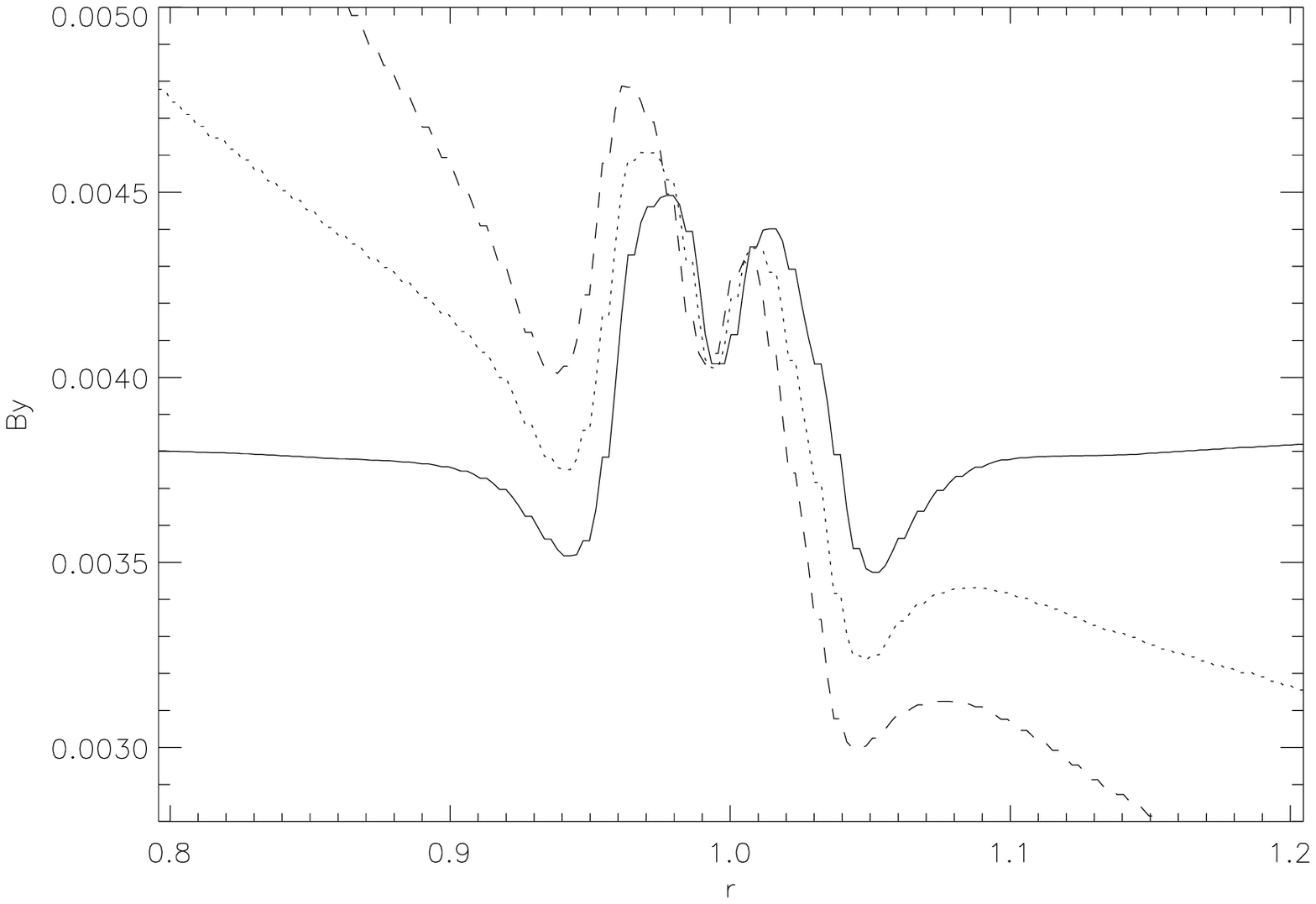} 
}
\caption[]{Radial profile of the toroidal component of the magnetic field at 
the azimuth of the planet for model G2 ({\it solid line}), G5 
({\it dotted line}) and G6 ({\it dashed line}).}
\label{fig9}
\end{figure}

For each of these models, the time history of the total torque exerted
by the disc on the planet is plotted in figure~\ref{fig8}. The solid
curve corresponds to model G1, while the other curves are deduced from
model G2 ({\it dotted line}), G5 ({\it dashed line}) and G6 ({\it
dotted--dashed line}). All of them were run for $160$~orbits, a long
enough time to reach saturation, which occurs between $80$ and $90$
orbits. The total torques obtained in models G1 and G2 are very
close. This is because the effects of the magnetic resonances located
on both sides of the orbit of the planet cancel each other in G2. The
hydrodynamic result is recovered in that case.  However, there is a
clear tendency for the torque to increase (ie becoming less negative/more 
positive) when the magnetic field
becomes steeper. As was already described above, it is negative for
model G2. For model G5, it is almost equal to zero, indicating that
the migration of the planet should be very slow in that case. Finally,
for model G6, it is positive. The planet should show strong outward 
migration.  This
trend can be qualitatively understood with the help of
figure~\ref{fig9}, where the radial profile of $B_{\phi}$ at the
azimuth of the planet is plotted for model G2 ({\it solid line}), G5
({\it dotted line}) and G6 ({\it dashed line}). As the magnetic field
gets steeper, the inner magnetic resonance gets stronger and the outer
magnetic resonance gets weaker. The balance between the positive
torque exerted in the neighborhood of the inner magnetic resonance and
the negative torque exerted near the outer magnetic resonance is biased in
favor of the former and the net torque increases.

We now compare these values of the torque with the values computed
from the linear analysis (see T03).  To recover SI units, the torques
given above have to be multiplied by $G M_{pl} M_\star / r_p$.  This
gives $T=-3.2 \times 10^{29}$, $5.34 \times 10^{28}$ and $4.5 \times
10^{29}$~SI for $p=0, 1$ and 2, respectively.  The linear analysis
gives $T=-5.06 \times 10^{29}$, $-1.01 \times 10^{29}$ and $3.7 \times
10^{29}$~SI for $p=0, 1$ and 2, respectively.  The agreement for $p=0$
and $p=2$ is very good, the difference between the torques being 37
and 21\%, respectively.  The agreement is not as good for $p=1$, but
note that in that case the torque is small, which makes it more
difficult to get a good accuracy from the simulations.  The torque
obtained from linear theory in the case $B=0$ is $-5.04 \times
10^{29}$~SI, very close to the torque corresponding to $p=0$.  This
also agrees very well with the numerical simulations.

\subsection{Migrating planets}  
\label{Migration}
In addition to performing simulations in which the planet is
held on a fixed circular orbit, runs were performed to investigate
the actual migration of the planet as a function of the magnetic
field profile in the disc. These runs are listed as N2, N3 and N4 in 
table~\ref{models}. 

\begin{figure}
\centerline{
\epsfig{width=12.cm, angle=0, file=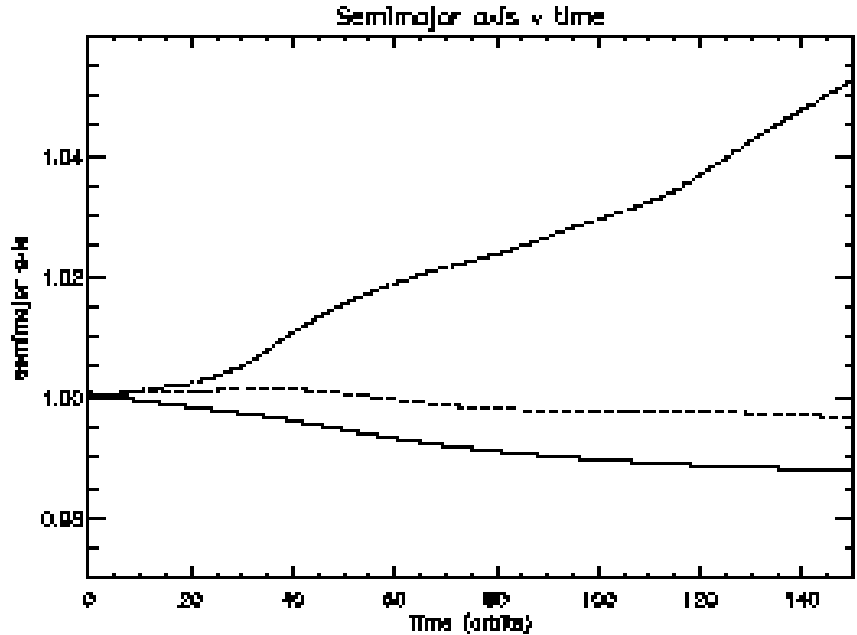}
}
\caption[]{Time evolution of the semimajor axes for simulations
N2 (solid line), N3 (dashed line) and N4 (dashed-dotted line).
While simulations N2 and N3 lead to inward migration, simulation
N4, with the larger magnetic field gradient, leads to outward migration.}
\label{fig10}
\end{figure}

The variation of semimajor axis with time is shown in
figure~\ref{fig10} for these three runs. Run N2 is shown by the solid
line, N3 by the dashed line, and N4 by the dotted-dashed line. These
calculations are in basic agreement with their GLOBAL counterparts,
and analytic expectations, in that they display quite rapid inward
migration (N2 and G2), slow migration (N3 and G5), and strong outward
migration (N4 and G6). These latter simulations provide definitive
evidence that a strong radial gradient in magnetic field in an
accretion disc can stop and even reverse type~I migration.  

\begin{figure*}
\epsfig{width=17.cm, angle=0, file=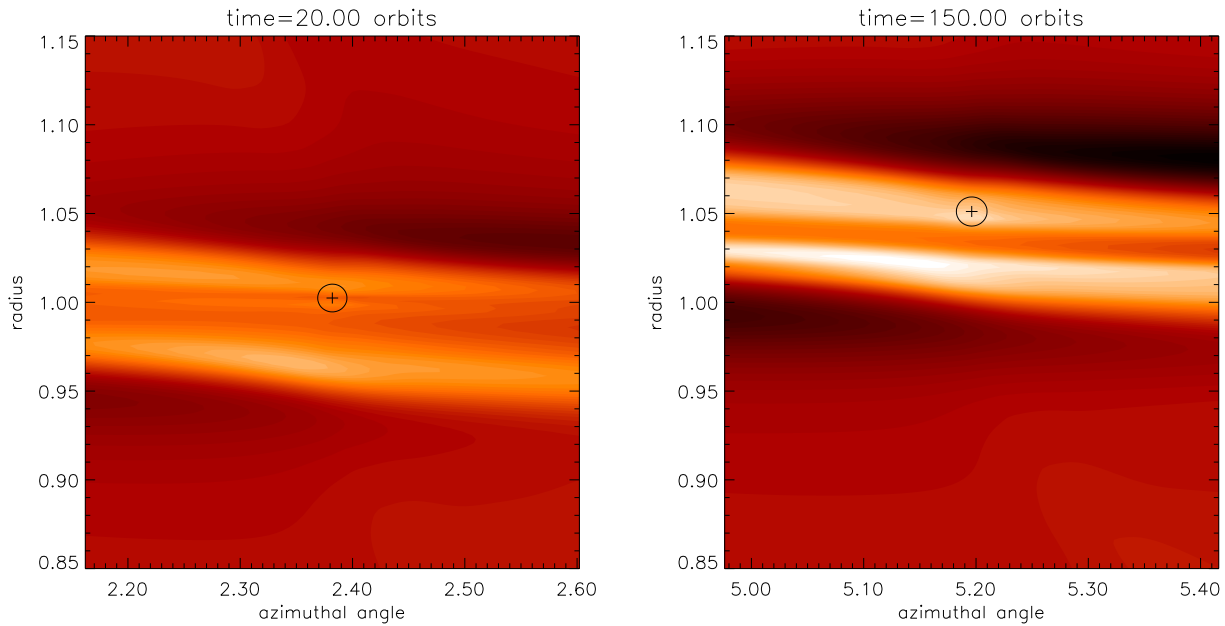}
\caption[]{The left panel shows a contour plot of $r^2 B_{\phi}$ for
run N4 at $t=20$. The right panel shows a similar plot
at time $t=150$. The locations of the magnetic resonances are
seen to track the radial location of the planet
(shown by the black cross) as it migrates outward.}
\label{fig11}
\end{figure*}

In figure~\ref{fig11} we present contour plots of $B_{\phi}$ in the
vicinity of the planet for run N4 at $t=20$ and $150$ orbits. The left
hand panel shows the magnetic resonances being established near the
beginning of the simulation. The right panel shows the field strength to have
increased at the resonances after 150 orbits, and also shows the
resonant locations tracking the radial position of the planet as it
migrates outward.  We note that the planet position (shown by a black
cross) appears to be closer to the outer magnetic resonance than the
inner one at this time. We speculate that this may be due to the
readjustment time of the disc being comparable to the migration time
in this run. Indeed, the disc response time is about 70--80
orbits, as shown by figure~\ref{fig5}, which is the time required for
the torques to approach steady values, and for the magnetic field to
achieve an approximate steady state at the magnetic resonances.  

In section~\ref{discmodel} we discussed briefly the inclusion of the
axisymmetric component of self--gravity in our simulations of
migrating planets.  Normally the disc mass used in simulations of the
type presented here is sufficiently small that this is not
required. However, because the simulations here are of high
resolution, and therefore computationally expensive, we are compelled
to use a fairly massive disc with correspondingly faster migration
rates. Not including the disc self--gravity, but including the
acceleration by the disc on the planet's orbital evolution, has the
effect of increasing the inward radial force experienced by the
planet, thus increasing its angular velocity.  This causes a shift in
the angular frequency associated with the planet's orbital motion that
modifies slightly the positions of the Lindblad and magnetic
resonances, making the outer resonances stronger relative to the inner
ones. Including the disc self--gravity helps to remove this
effect. Given the close proximity of the magnetic resonances to the
planet orbital radius, we have found that such shifts in frequency can
qualitatively change results. For example, we ran a case similar to
run N4, but without the disc self--gravity, and found the outward
migration stalled after $t < 100$ orbits.  Including the self--gravity
removes this stalling effect.

\section{Discussion}

We have performed numerical simulations of a low--mass planet embedded
in a disc containing a toroidal magnetic field using two different
codes.  In the runs performed with GLOBAL, the grid was logarithmic and
the planet was kept on a fixed circular orbit.  By contrast, in the
runs performed with NIRVANA, the grid was uniform and the planet
was allowed to migrate under the effect of the torque exerted by the
disc.

An important aspect of these simulations was to study the effect of the
magnetic resonances on the migration of the planet.  Since these
resonances are located very close to the planet's orbit, at only a
fraction of the disc semi--thickness, a very high resolution was
required.

In agreement with the linear theory, we have found that
magneto--acoustic waves propagate radially away from the planet beyond
the Lindblad resonances, whereas slow--MHD waves propagate essentially
along the field lines at the location of the magnetic resonances.  For
the parameters we adopted in this paper, the torque exerted by the
region of the disc around these resonances tended to dominate the
Lindblad torque.  We found inward and outward migration for a uniform
field and for a field varying as $r^{-2}$, respectively.  The torques
computed from the simulations were in very good agreement with those
calculated from the linear theory.  For a field varying like $r^{-1}$,
we found outward migration with GLOBAL and inward migration with NIRVANA, 
in both cases at a rate much reduced compared to when the
field is uniform.  The linear theory predicted inward migration in
that case, also at a reduced rate.  This shows that the
simulations performed here are at the limit of what can be achieved
today.

These simulations confirm that, if the magnetic field has a gradient
which is {\em locally} steep enough, migration of low--mass planets
can be reversed.

An interesting feature that was seen in the simulations is that, if
the adjustement time of the disc (and therefore of the resonances) is
comparable or longer than the migration time, which may be the case
for a massive disc, then the position of the planet with respect to
the resonances is affected.  In the case $p=2$, the planet, which was
migrating outward, got closer to the outer resonance.  We may then
expect the migration to be slowed down, as the outer resonance becomes
more important, until the disc readjusts.  By contrast, for a planet
migrating inward, the inner resonance would get closer and in that
case again migration would slow down.

Note that the slow MHD waves that propagate around the magnetic resonances are
the modes which are destabilized in the magnetorotational instability
(Balbus and Hawley 1991).  Therefore, it is not clear that the
processes that are described here would still operate if the field
were weak enough to be unstable.  Note however that the modes that are
magnetorotationally unstable have large values of $m$ (Balbus and
Hawley 1992, Terquem and Papaloizou 1996), whereas the modes that are
important here have low values of $m$.  So it is possible that the
resonances described here still operate to reverse planet migration in
an unstable disc, i.e. in the presence of turbulence.

In the simulations presented in this paper, as in the analysis of T03,
the disc is assumed to be infinitessimally thin.  As the distance between
the planet and the resonances involved in the disc/planet interactions is
smaller than the disc semithickness, this approximation is actually not
valid.  However, in the absence of a magnetic field, results from 3D
calculations were not found to be qualitatively different from those from
2D calculations.  This is because the total Lindblad and corotation
torques of 3D waves (with vertical motions) are much smaller than those
of 2D waves.  So the only effect of the 2D approximation is to introduce
a vertical averaging of the gravitational potential of the protoplanet. 
This results in the Lindblad and corotation torques being 60\% and 50\%
larger, respectively, in 2D than in 3D (Tanaka et al. 2002).    In the
case of a magnetized disc, we anticipate that the torque from the
magnetic resonances will be similarly reduced in 3D.

It is the angular momentum carried by the perturbation in the vicinity
of the magnetic resonances which enables the planet to reverse its
migration.  When there is no magnetic field, the angular momentum
carried by the density waves beyond the Lindblad resonances is carried
away from the planet.  The waves that propagate around the magnetic
resonances are trapped there though.  So it is not clear how the
angular momentum they carry is transported away.  If the waves were
just bouncing back and forth in some finite region, there would be no
net exchange of angular momentum between the planet's orbital motion
and the disc's rotation.  Therefore, it is possible that the modes
that propagate around the magnetic resonances tunnel through the
region where the disc response is evanescent.  The detail of the
processes by which angular momentum is exchanged will be the subject
of a future paper.

\section*{acknowledgement}
This work was supported in part by the European Community's Research
Training Networks Programme under contract HPRN--CT--2002--00308,
``PLANETS''.  CT acknowledges partial support from the Institut
Universitaire de France.  

The numerical simulations presented in this paper were performed on the 
U.K. Astrophysical Fluid Facilities (UKAFF) and on the Queen Mary University 
of London HPC cluster purchased under the SRIF initiative.

\newpage

\appendix
\section{The logarithmic grid}

In this appendix, we describe in detail the logarithmic grid that was used 
for model G1 to G6. Its aim is to increase the resolution in the 
neighborhood of the planet with a low cost on the computing time.

The radial grid is composed of three zones: first, a buffer zone, 
$r \in [Rb1,Rb2]$, where each cell has the same size, encompasses the orbit 
of the planet. In the region $[R_{in},Rb1]$, the resolution increases has 
$r$ increases: $(\Delta r)_{i}/(\Delta r)_{i+1}=h$. This is the opposite in 
the region $[Rb2,R_{out}]$. There, again, the size of the cells increases as 
one goes away from the planet. 

The azimuthal grid has similar properties. Since the planet is located at 
$\phi_{pl}=\pi$, there is a buffer zone, with constant grid spacing, in the 
interval $\phi \in [\pi-\pi/20,\pi+\pi/20]$. On both sides of this 
zone, the size of the grid cells is increasing with the same ratio $h$ as 
for the radial grid.

\begin{figure}
\centerline{
\epsfig{width=12.cm, angle=0, file=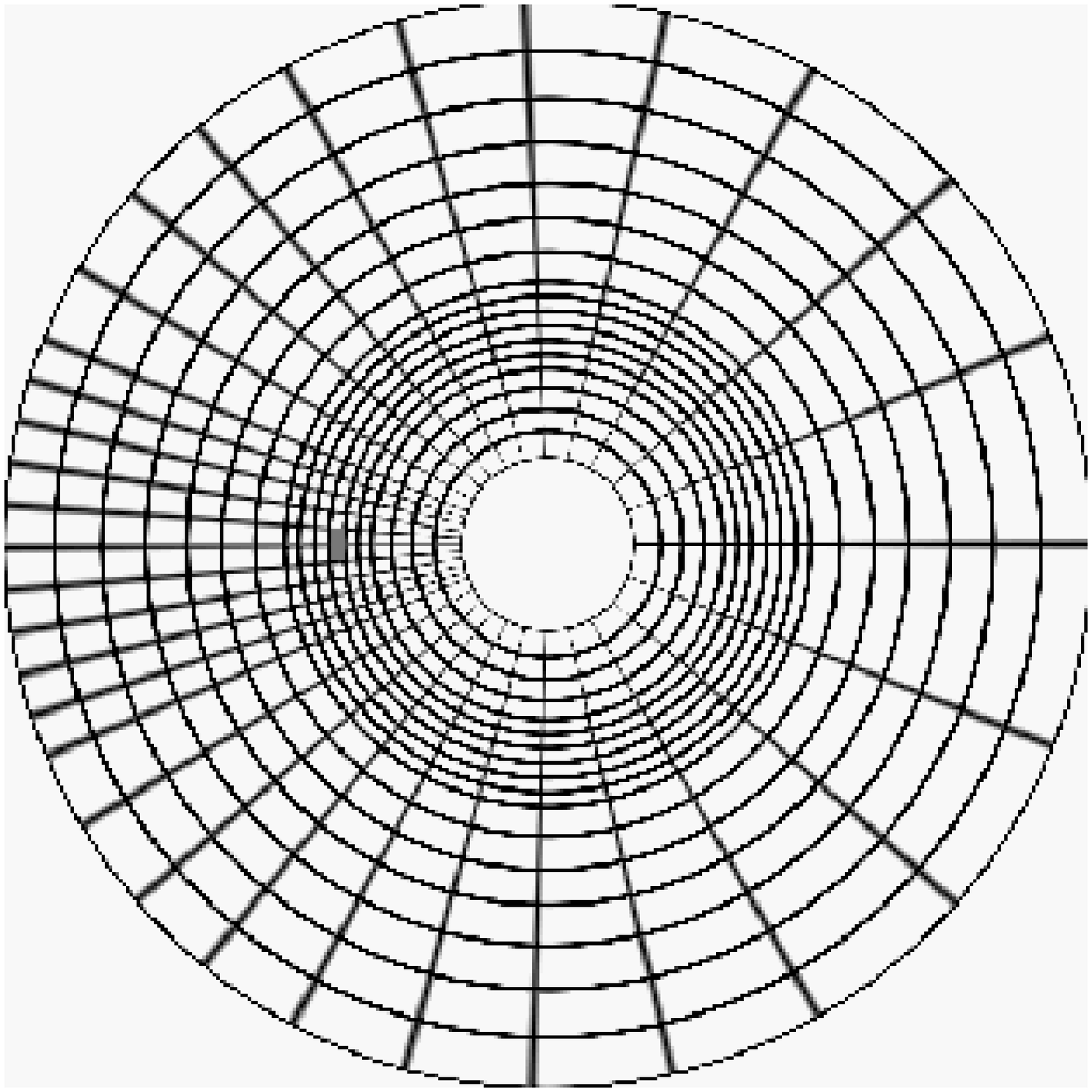} 
}
\caption[]{Cartoon illustrating the properties of the grid. The position of 
the planet is shown by the grey area. The resolution is higher there than 
what it is elsewhere in the computational domain. Note that the ratio between 
neighboring cells sizes has been exaggerated here for clarity.}
\label{fig grid}
\end{figure}

Figure~\ref{fig grid} illustrates the resulting grid topology. The planet is 
located in the shaded grid zone, where the resolution is the highest. Note 
that the ratio between neighboring cells sizes has been exaggerated for 
clarity in this figure. For the actual simulations presented in this paper, 
we used $Rb1=0.9$, $Rb2=1.1$ and $h=1.02$. For our fiducial run, model G2, we 
used $60$ cells in the buffer zone of the radial grid, which 
gives a resolution $\Delta r=3.3 \times 10^{-3}$. The cell size in 
$\phi$, which is not so crucial for a good description of the magnetic 
resonances, was taken to be twice this value at the location of the planet: 
$r_{pl}\Delta\phi=6.6 \times 10^{-3}$. On a uniform grid, the number of cells 
required to reach such a resolution would have been $N_r=630$ and 
$N_{\phi}=950$. Here, with the help of the logarithmic grid, the actual 
resolution is $(N_r,N_{\phi})=(242,258)$. For model G3, the number of 
cells in the buffer zone was doubled in the azimuthal direction, while for 
model G4 it was composed of $80$ cells in both directions.

\begin{thebibliography}{} 

\bibitem[1991]{Balbus1}
Balbus S.~A., \& Hawley J.~F., 1991, ApJ, 376, 214

\bibitem[1992]{Balbus2}
Balbus S.~A., \& Hawley J.~F., 1992, ApJ, 400, 610

\bibitem[1998]{Balbus3}
Balbus S.~A., \& Hawley J.~F., 1998, Rev. Mod. Phys., 70, 1

\bibitem[1998]{Boss1} 
Boss A.~P., 1998, ApJ, 503, 923
  
\bibitem[2004]{Fromang}
Fromang S., 2004, PhD Thesis

\bibitem[1979]{Goldreich1}
Goldreich P., \& Tremaine S., 1979, ApJ, 233, 857 (GT79)

\bibitem[1995]{Hawley0}
Hawley J.~F., Stone J.~M., 1995, Comput. Phys. Commun., 1995, 89, 127

\bibitem[2003]{Lecar} 
Lecar M., Sasselov D. D., 2003, ApJ, 596, 99

\bibitem[1993]{Lin} 
Lin D.~N.~C., \& Papaloizou J.~C.~B., 1993, in Protostars and
Planets III, eds. E.~H. Levy, \& J.~I. Lunine (Tucson: University
of Arizona Press), p.~749

\bibitem[2003]{Matsuyama}
Matsuyama I., Johnstone D., Murray, N., 2003, ApJ, 585, 143

\bibitem[2000]{Nelson1}
Nelson, R.~P., Papaloizou J.~C.~B., Masset F., Kley W., 2000, MNRAS, 318, 18

\bibitem[2004]{Nelson2}
Nelson, R.~P., Papaloizou J.~C.~B., 2004, MNRAS, 350, 849

\bibitem[1996]{Pollacketal}
Pollack J.B., Hubickyj O., Bodenheimer P.; Lissauer J.~J., Podolak M., Greenzweig Y., 1996, Icarus, 124, 62

\bibitem[2002]{Tanakaetal}
Tanaka H., Takeuchi T. \& Ward W.~R., 2002, ApJ, 565, 1257

\bibitem[1996]{Terquem}
Terquem C., \& Papaloizou J.~C.~B., 1996, MNRAS, 279, 767

\bibitem[2003]{Terquem1} 
Terquem C., 2003, MNRAS, 341, 1157 (T03)

\bibitem[1998]{Trilling} 
Trilling D.~E., Benz W., Guillot T., Lunine J.~I., Hubbard W.~B., Burrows A., 1998, ApJ, 500, 428

\bibitem[2003]{Udry1}
Udry S., Mayor M. \& Santos N.~C., 2003, A\&A, 407, 369

\bibitem[1986]{Ward1}
Ward W.~R., 1986, Icarus, 67, 164

\bibitem[1997]{Ward3}
Ward W.~R., 1997, Icarus, 126, 261

\bibitem[1997]{Ziegler} 
Ziegler U., Yorke H.~W., 1997, Comput. Phys. Commun., 101, 54

\end{thebibliography}
\end{document}